%% file: ycguo.cen.tex
\documentclass[usenatbib]{mn2e}
\usepackage{graphicx,times}

\input{macros}

\title[Structural Properties of Central Galaxies]{Structural Properties of Central Galaxies in Groups and Clusters}
\author[Guo et al.]
{\parbox[t]{\textwidth}{Yicheng Guo$^{1}$\thanks{E-mail: yicheng@astro.umass.edu}, 
Daniel H. McIntosh$^{1}$\thanks{Current address: Department of Physics, University of Missouri-Kansas City, Kansas City, Missouri 64110, USA}, 
H. J. Mo$^{1}$, Neal Katz$^{1}$, Frank C. van den Bosch$^{2}$, \\
Martin Weinberg$^{1}$, Simone M. Weinmann$^{3}$, Anna Pasquali$^{2}$, Xiaohu Yang$^{4,5}$}\\
\vspace*{3pt} \\
$^1$ Astronomy Department, University of Massachusetts,
710 N. Pleasant St., Amherst, MA 01003, USA\\
$^2$ Max-Planck-Institut f\"ur Astronomie,
K\"onigstuhl 17, D-69117 Heidelberg, Germany\\
$^3$ Max-Planck-Institut f\"ur Astrophysik, Karl Schwarzschild Str. 1, 
Postfach 1317, 85741 Garching, Germany\\
$^4$ Shanghai Astronomical Observatory, the Partner Group of MPA,
Nandan Road 80, Shanghai 200030, China\\
$^5$ Joint Institute for Galaxy and Cosmology of Shanghai Astronomical
Observatory and University of Science and Technology of China}

\begin{document}

\date{{\sc Draft: } \today }

\pagerange{\pageref{firstpage}--\pageref{lastpage}} \pubyear{2008}

\maketitle

\label{firstpage}

\begin{abstract}
Using a statistically representative sample of 911 central galaxies (CENs) from the SDSS
DR4 Group Catalogue, we study how the structure (shape and size) 
of the first rank (by stellar mass) group and cluster members depends on (1) galaxy stellar mass
(${\rm M_{star}}$), (2) the global environment defined by the dark matter halo mass (${\rm M_{halo}}$)
of the host group, and (3) the local environment defined by their special 
halo-centric position. 
We establish a GALFIT-based pipeline for 2D \sersic fitting of SDSS data
to measure the \sersic index, $n$, and half-light radius, $r_{50}$, 
from {\it r}-band galaxy images.
Through tests with simulated and real image data, 
we demonstrate that our pipeline can recover galaxy properties without 
significant bias. We also find that uncertainties in the 
background sky level translate into a strong covariance between the
total magnitude, the half-light
radius, and the \sersic index, especially for bright/massive galaxies.
We apply our pipeline to the CEN sample and
find that the \sersic index $n$ of CENs depends strongly on ${\rm M_{star}}$, 
but only weakly or not at all on ${\rm M_{halo}}$. 
The $n$-\mstar relation
holds for CENs over the full range of halo masses that we consider.
Less massive CENs tend to be disk-like and high-mass systems are typically
spheroids, with a considerable scatter in $n$ at all galaxy masses.
Similarly, CEN sizes depend on galaxy stellar mass and luminosity, 
with early and late-type galaxies exhibiting different slopes for
the size-luminosity ($r_{50}$-$L$) and the size-stellar mass 
($r_{50}$-${\rm M_{star}}$) scaling relations.
Moreover, to test the impact of local environment on CENs, we compare the structure of CENs with that of comparable 
satellite galaxies (SAT).
We find that low mass ($<10^{10.75}{\rm h^{-2}M_{\odot}}$) SATs have somewhat
larger median \sersic indices compared with CENs of a similar stellar mass.
Also, low mass, late-type SATs are moderately smaller in size than 
late-type CENs of the same stellar mass. However, we find no size differences 
between early-type CENs and SATs and {\it no structural differences} 
between CENs and SATs when they are {\it matched in both optical colour and
stellar mass.}
The similarity in the structure of massive SATs and CENs
demonstrates that this distinction has no significant
impact on the structure of spheroids.
We conclude that \mstar is the most fundamental property determining the 
basic structural shape and size of a galaxy.
In contrast, the lack of a significant $n$-\mhalo relation rules out
a clear distinct group mass for producing spheroids, and 
the morphological transformation processes that produce
spheroids must occur at the centres of groups spanning a wide range of masses.
\end{abstract}


\section{Introduction}
\label{intro}

Understanding the role that environment plays in shaping the morphology of galaxies remains an 
important challenge in the study of galaxy formation and evolution. 
In the standard galaxy formation and evolution paradigm, all galaxies started as
star-forming disks at the centres of small dark matter (DM) haloes.
Subsequent hierarchical evolution has transformed galaxies into
spheroids to produce the bimodality observed in the present-day populations.
Tied to the strong colour bimodality are differences in morphology
and related galactic structure. Blue star-forming galaxies
tend to be disk-dominated with exponential radial light profiles
(late-types), while red non-star-forming systems have spheroidal
mass distributions with steeper light profiles (early-types).
Many observations suggest that there may be an environmental
component to differences in galaxy properties; e.g., dependence of
morphology \citep[e.g.,][]{dressler80a,goto03,mcintosh04a,blanton05a} and star formation related
properties \citep[e.g.,][]{hashimoto98,balogh04a,kauffmann04} to local galaxy density. Presumably, the physical processes that have built
larger DM haloes housing groups and clusters of galaxies
must be responsible at some level for the observed evolution of the two 
primary galaxy populations. 

The latest theoretical models of galaxy evolution
invoke the transformation of blue disks into red spheroids
within a hierarchical framework to explain the factor of two
growth observed in the early-type galaxy (ETG) population
since $z=1$ \citep[e.g.,][]{bell04b,blanton06,borch06,faber07,brown07}.
It is usually assumed that to successfully reproduce the
galaxy bimodality requires physical mechanisms that {\it both transform
star formation and morphology}. 
A variety of galaxy transformation scenarios
can be found in the literature,
many of which are predicted to be important only in particular
environments. 
Yet, recent advances in understanding bimodality have
focused on environmental processes that mostly impact star formation.
For example, \citet{vandenbosch08a} successfully demonstrate that
quenching is important for producing redder galaxies over a range of
different environments,
but a clear picture of what governs morphological bimodality
is still lacking and controversial.
To shed light on whether a 
special environment exists for the transformation of galaxy morphology, we 
study the 
shapes and sizes of a representative sample of central galaxies from galaxy 
groups 
and clusters in the Sloan Digital Sky Survey \citep[SDSS,][]{york00}.
Quantifying the structural properties of galaxies, such as the
steepness of the light profile shape,
provides a direct means to assess morphological transformation.

If all galaxies started as small disks, then the simple existence
of spheroidal systems makes it clear that morphological transformation
occurs.
The transformation from late to early-type galaxies may occur in one
traumatic event or be the result of multiple processes over billions
of years.
There is no shortage of theoretical predictions for the creation
of ETGs, from violent galaxy-galaxy mergers to the slow
fading of disks. Generally speaking, the early type population includes a
range of spheroid-dominated morphologies with and without a disk
component (e.g., ellipticals, lenticulars, and
bulge-dominated spiral galaxies). 
Thus, it is likely that more than one
physical mechanism is responsible for the variety among ETGs.
Narrowing
down the processes that are most responsible for turning disks into
spheroids remains a critical piece of the galaxy evolution puzzle.
In this paper, we approach this problem by testing whether or not
there is a specific environment where a strong morphological transition
takes place that can then be tied to a particular process.

There are several morphologically-altering mechanisms that are predicted
to be effective mainly in the high-density environs of massive
groups and clusters.
Harassment, the cumulative effect of many high speed encounters with satellite 
galaxies \citep{moore96a}, is predicted to occur primarily in groups and
clusters after a satellite galaxy is accreted, and may transform a 
disk galaxy into a more early-type morphology by heating and 'puffing up'
the disk component.
Tidal stripping, 
the effect of the tidal force suffered by a satellite along its orbit, 
may transform a satellite galaxy into a spheroid by removing its disk,
and it may be effective in haloes over a large mass range.
Since the above two processes 
change the stellar mass of a galaxy by at most a factor of two, if they are
responsible for 
the morphological transformation one would expect to see satellites with
statistically significant differences in their morphological and structural
properties compared with centrals of a similar stellar mass. 
However, \citet{vandenbosch08a} and \citet{weinmann08}
find that satellite galaxies are only marginally more concentrated than
central galaxies with same stellar mass, which suggests that satellite galaxies 
only undergo a minor change in their morphology after 
they fall into a massive cluster, and that the above two processes may not 
be sufficient to explaining their major morphological transformation.
It is important to note that these processes basically produce early types
by diminishing the disk structure of later-type spirals; i.e., these
mechanisms may produce S0/Sa early types, but they do not create massive 
spheroids or elliptical galaxies.

Besides the aforementioned cluster-specific mechanisms that appear to have only minor
impact on the morphologies of satellite galaxies, 
numerical simulations demonstrate that the merger of two disk galaxies 
with similar masses produces a spheroid
galaxy \citep[e.g.,][]{toomre77,barnes96,naab03,cox06} and, hence, are
likely an important mechanism for the formation of spheroids and ellipticals. 
Major mergers are believed to be 
efficient in group-size haloes and to be suppressed in more massive haloes
because of the increasing differences in the relative velocities between member 
galaxies compared to their own internal velocity dispersions.
It has long been assumed that the smaller velocity dispersions
found in galaxy groups allow
more galaxy interactions \citep{cavaliere92}; also 
the orbital decay timescale is shorter in
lower-mass haloes \citep[e.g.][]{cooray05d}.
However, \citet{mcintosh08} find
that major mergers among 
massive galaxies occurs at the centres of clusters, as well as large
groups, yet this 
merely makes bigger spheroids from smaller spheroids. 
Moreover, \citet{hopkins08} also show,
using theoretical arguments, that the
massive central galaxies of clusters still have a large chance 
to merge with their satellites. These findings highlight once again
the open question of which environment or halo mass 
is ideal for transforming disks into spheroids. It appears
that major mergers preferentially happen between central and satellite galaxies;
therefore, the existence of a special halo mass for major morphological
transformations such as mergers might manifest itself as a noticeable
change in the morphological distribution of central galaxies at some
specific halo mass. We already know that central
galaxies living in small haloes have disk-like shapes with low 
concentrations and flat light profiles while those in large haloes have 
spheroid-like shapes with high concentrations and steep light profiles.
A careful study of
the distribution of structural properties in haloes spanning a range in
mass will help shed light on this open question.

Our analysis is based on two quantitative
measures of galactic structure, the \sersic index and the size, which
are directly related to galaxy morphology.
The measurement of galaxy light profile shapes and sizes has a long
history \citep[][and references therein]{shaw89,byun95,dejong96c,simard98,
khosroshahi00} and the development of automatic routines to handle
the huge number of galaxies from modern surveys is well-motivated.
In this work, we develop a powerful pipeline
for applying a well-tested and popular software package for two-dimensional
(2D)
galaxy image fitting \citep[GALFIT,][]{peng02} to SDSS $r$-band data.
We fit a \sersic model \citep{sersic68} to the images of galaxies and use the 
\sersic index, $n$, and half-light radius, $r_{50}$, of the best fit models
to describe the shapes and sizes of galaxies, respectively.
A well-known, yet often overlooked, issue in profile fitting
is the critical sensitivity to the estimate of the
background sky level \citep{macarthur03}. We explore this issue in detail
and discuss its impact on quantifying the structure of galaxies, in
particular that of high-mass central galaxies. 

In this paper, we try to answer two questions about the environment of morphological 
transformation: (1) is there a critical DM halo mass
where central galaxies 
are transformed from late-type to early-type? (2) is the central
position in groups and 
clusters a special place for determining the structure of galaxies.
Previous work arguing whether galaxy morphology
 \citep[e.g.,][]{dressler80a} or star formation \citep[e.g.,][]{vanderwel08}
depends more critically on environment employ local galaxy density
measurements that are less natural and less physically meaningful than
the halo mass and the location within a host halo \citep{weinmann06a}.
Here we use the SDSS DR4 Group Catalogue of \citet{yang07}, which provides
statistical measures of the host halo mass (global environment) 
and the halo-centric position (local environment) for SDSS galaxies.
We study the quantitative structure
of galaxies living at the presumed dynamical centres of DM haloes spanning
nearly three orders of magnitude in mass. 
In hierarchical models of galaxy formation, all galaxies begin as 
the central galaxy in a smaller halo and then become satellite galaxies if the 
small halo merges with a bigger one. 
Therefore, comparing the structural properties of
centrals and satellites can shed light on the 
evolution of galactic morphology and its dependence on environment,
and help answer whether central galaxies are a distinct population with 
unique formation histories.

In \S\ref{sample} we present our sample selection. In \S\ref{pipeline}
we present and
test our fitting pipeline and show the sensitivity of the fits to the
background sky level. We present the results of our fits to the central
galaxies in \S\ref{results},
both their \sersic indices and sizes, and compare them to satellite
galaxies.  We summarise our main conclusions in \S\ref{summary}. 
We also include an appendix where we compare our fits to those of the NYU-VAGC
\citep{blanton05nyu}.
Throughout we adopt a flat ${\rm \Lambda CDM}$ cosmology with 
$\Omega_m=0.3$, $\Omega_{\Lambda}=0.7$ and use the Hubble 
constant in terms of $h\equiv H_0/100 {\rm km~s^{-1}~Mpc^{-1}}$.

\section{Sample Selection}
\label{sample}

\subsection{Central Galaxy Sample from SDSS Group Catalog}
\label{groupcat}

To study the structural properties of central galaxies (CENs) in groups and clusters, 
we use the SDSS galaxy group catalogue
of \citet[hereafter Y07]{yang07}. This catalogue is 
constructed using the New York University Value-Added Galaxy Catalog \citep[NYU-VAGC,][]{blanton05nyu} reprocessing of the spectroscopic 'Main' selection \citep{strauss02} from 
the fourth data release of Sloan Digital Sky Survey \citep[SDSS DR4,][]{adelman05}. The NYU-VAGC provides improved reductions and additional galaxy property measurements. 
The group catalogue
provides two physically motivated measures of environment for each galaxy:
(i) the dark matter halo mass of its host group (${\rm M_{halo}}$), and (ii) 
its distance to the central (i.e, highest stellar mass) group member of the group. Here we briefly describe the group catalogue 
and we discuss
its redshift completeness as it relates to our final selection of a 
representative sample of CENs.

The details of the group finder used to construct the DR4 version of the group 
catalogue are 
given in \citet{yang05a,yang07}. Briefly, the group 
finder starts with a friends-of-friends (FOF) algorithm using a restrictively small linking length 
to define potential groups and their centres in a galaxy redshift survey. A rough group mass is estimated from the total group luminosity, assuming a mass-to-light ratio\footnote[1]{The
resulting  group catalogue  is insensitive  to the  initial assumption
regarding the M/L ratios.}. From the group mass, the group-finder uses an adaptive filter to iterate on the virial search parameters (projected radius and velocity
dispersion), which in turn are used  to select group members
in redshift  space.  This method  is iterated until the  group members
converge.  This algorithm has been thoroughly tested with mock galaxy redshift surveys and 
is shown to be more successful than conventional FOF finders. The
average completeness  of individual groups in terms of membership is $\sim 90$
percent, with  only $\sim 20$ percent contamination from interlopers.
The halo mass of each group is estimated 
using two methods:
(1) the ranking of its total characteristic luminosity and (2)
the ranking of its total characteristic stellar mass. As shown
in Y07, both methods agree very well with each other, with a scatter of 0.1 
dex (0.05 dex) at low (high) halo masses. In this paper, we use 
the halo mass from stellar mass ranking. Finally, owing to the $r<17.7$ (extinction-corrected)
magnitude limit of the Main sample, the minimum \mhalo for which the group selection is 
complete changes with redshift. 
Therefore, we limit our analysis to groups with $z\leq0.08$, which are detected with high completeness down to ${\rm log(M_{halo}/h^{-1}M_{\odot})}=11.78$, allowing us to study galaxies 
from small groups with good image resolution.

The spectroscopic completeness of the sample used to determine galaxy groups
plays a crucial role in identifying CENs.
Owing to fibre collisions, the Main sample misses about 8 percent of SDSS
galaxies meeting the spectroscopic targeting criteria \citep{blanton03d}. 
This effect is severe in regions of high galaxy number density
\citep{hogg04}, such as in large groups and clusters.
The Y07 group catalogue contains three samples that address this incompleteness 
differently. Each sample spans the redshift range of $0.01\leq z\leq0.20$. Sample I 
contains 362356 Main galaxies. Sample II includes 7091 additional galaxies with 
spectroscopic redshifts measured from other surveys 
\citep[e.g., 2dF,][]{colless01}. Sample III adds 
38672 galaxies missing redshifts 
due to fibre-collisions, which are assigned a redshift based on their nearest neighbour.
 For the three samples, the group finder detects 201621, 204813 and 205846 groups, 
respectively. 

Note that the fibre-collision correction applied to sample III may transfer
some CENs in sample II to satellite galaxies in sample III and vice versa. 
For our selection, we choose only galaxies identified as CENs in both
sample II and sample III. These galaxies are guaranteed to be 
CENs regardless of whether the fibre-collision correction was 
applied or not. In addition, we use the
\mhalo of these CENs drawn from sample II.
Our analysis (discussed in \S\ref{pipeline}) is based on a CPU-intensive galaxy image 
fitting routine, thus, to construct a representative sample we randomly 
select CENs from halo mass bins spanning the full range of groups
in the $z \leq 0.08$ volume-limited sample. 
For ${\rm log(M_{halo}/h^{-1}M_{\odot})}=[12.0:14.0]$, we randomly select 100 CENs from eight bins of 0.25 dex
width. At higher halo masses the number of groups decreases rapidly,
thus to maintain good statistics we select 100
CENs from the [14.0:14.5] bin, and use all 11 CENs in groups with 
${\rm log(M_{halo}/h^{-1}M_{\odot})}>14.5$.
The total number of CENs in our representative sample is 911. 

\begin{figure*}
\center{\includegraphics[scale=0.9, angle=0]{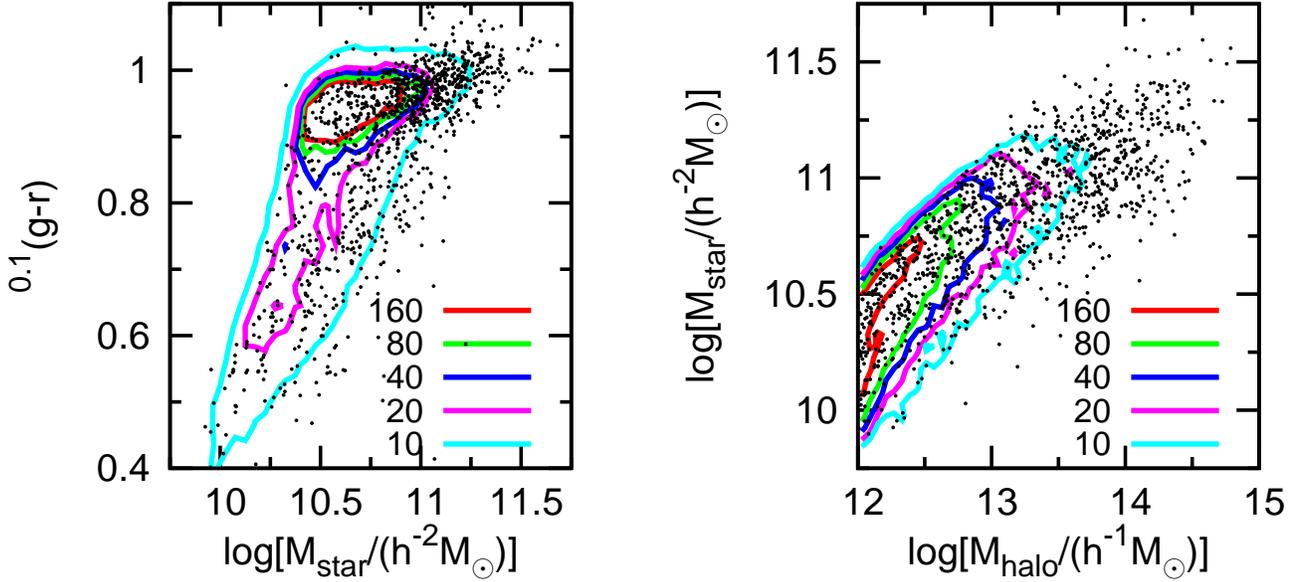}}
\caption[]{The colour-\mstar and ${\rm M_{star}}$-\mhalo distributions of our 
selected CEN sample (black circles). Contours show the
distribution of all $z\leq0.08$ CENs from the SDSS DR4 group catalogue. 
Contours with different colours connect bins with same number of galaxies as 
labelled. Each bin has a width of 0.03 in colour and 
0.05 dex in ${\rm M_{star}}$ in the {\it left} panel, and a width of 
0.05 dex in ${\rm M_{star}}$ and 0.06 dex in ${\rm M_{halo}}$ in the {\it 
right} panel.}
\label{fig:samplecheck}
\vspace{-0.2cm}
\end{figure*}

In Figure \ref{fig:samplecheck}, we plot the stellar mass (${\rm M}_{\rm star}$)
distributions of our CEN sample as a function of galaxy colour and halo mass.
For each galaxy, the NYU-VAGC provides the Petrosian $^{0.1}(g-r)$ colour,
K+E corrected to $z=0.1$ (see Y07 for details), 
and we estimate ${\rm M}_{\rm star}$ using the 
colour-derived M/L from \citet{bell03b}.
As shown in Figure \ref{fig:samplecheck}, our representative CEN selection
samples the same ${\rm M_{star}}$, \mhalo and $^{0.1}(g-r)$ space as all
CENs with $z\leq0.08$ in the group catalogue (shown by the contours).
We note that our selection of CENs
spans a wide range of \mstar running from 
$10^{9.8}$ to $10^{11.7}{\rm h^{-2}M}_{\odot}$. 
Owing to the exponentially decreasing number density of high-mass haloes and the 
${\rm M_{star}}$-\mhalo relation of CENs (Fig.1, right panel), 
our selection of a constant number of groups
per \mhalo bin results in a CEN sample that is heavily-weighted towards
higher stellar masses and redder colours.

\subsection{Matched Satellite Galaxy Samples}
\label{matching}

In addition to studying the structural properties of CENs, we also want to investigate whether 
their central position in groups or clusters produces a distinctive structural 
difference compared to non-CEN galaxies. 
Following a similar method as in \citet{vandenbosch08a},
we construct two control samples of satellite (SAT) 
galaxies, one to match CENs in stellar mass only, and the other to match in
both stellar mass and colour.  
For each CEN in our sample, we first randomly select from all SATs with 
$z\leq0.08$ a similar-mass counterpart with a ${\rm M_{star}}$
within $\pm0.08$ dex, hereafter the SAT sample S1. We next find
SATs likewise again matched to CENs in ${\rm M_{star}}$, but further
restricted to match their colour within $\Delta ^{0.1}(g-r) = \pm 0.03$,
hereafter the SAT sample S2.
Once a SAT is matched to a CEN, we remove it 
from the pool so that there is no duplication in each SAT sample.
The matching criteria are equal to the measurement uncertainties of the
stellar mass and the colour \citep{bell03b},
and using a matching criteria with either a
smaller ${\rm \Delta M_{star}}$ or $\Delta ^{0.1}(g-r)$ will greatly reduce 
the number of matched SATs, especially towards the massive end, 
because we use a volume-limited sample and don't allow duplicate
matches. For CENs with ${\rm log(M_{halo}/h^{-1}M_{\odot})}<11.0$ we find 
a matched SAT using the above criteria more than 90 percent of the time. 
At higher masses (${\rm log(M_{halo}/h^{-1}M_{\odot})}>11.3$), SATs
become rare and the fraction of massive CENs with a matched SAT rapidly 
drops to less than 10 percent. 
Hence, we achieve SAT samples of 769 (S1) and 746 (S2) matches.

\begin{figure*}
\center{\includegraphics[scale=0.8, angle=0]{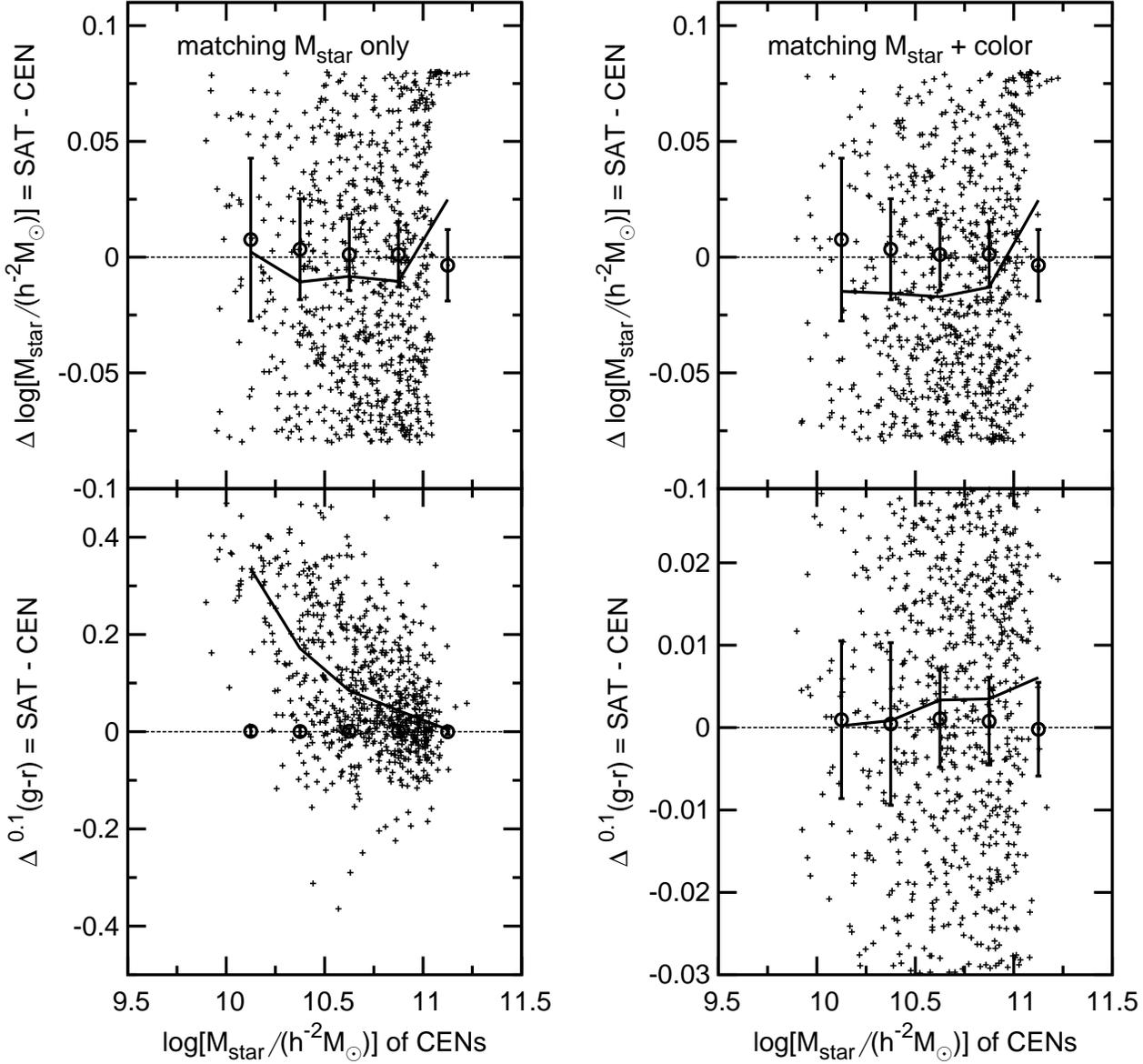}}
\caption[]{The difference of \mstar (\textit{upper} panels) and colour 
(\textit{lower} panels) between matched CENs and SATs as a function of 
CEN ${\rm M_{star}}$.
The \textit{left} column shows the differences for the SAT sample S1, which
is matched with our CEN sample only in stellar mass,
while the \textit{right} column shows the SAT sample S2,
which is matched with our CEN sample in both stellar mass and colour. 
The solid line in each panel shows the median of the difference for each
0.25 dex wide ${\rm M_{star}}$ bin. For comparison, circles with errorbars 
show the mean and 3$\sigma$
deviation of the biases in each mass bin from matching the CEN sample to 20 mock samples created from itself with random changes according to measurement uncertainties (see text for details).
\label{fig:matching}}
\end{figure*}

In Figure \ref{fig:matching}, we plot
the stellar mass and colour comparisons for our CEN-SAT matched samples as
a function of CEN stellar mass.
There is a slight bias in that SATs in both sample S1 and S2 are
less massive than their counterpart CENs with a median difference (SAT - CEN)
of -0.01 dex, 
and SATs in S1 are obviously 
redder than their counterpart CENs. 
The difference of colour for the SAT sample S2
({\it lower right} panel) is also small, with 
an almost zero median for low mass (${\rm log(M_{star}/h^{-2}M_{\odot})}<10.5$)
SAT-CEN pairs and a median of 0.005 for massive (${\rm log(M_{star}/h^{-2}M_{\odot})}>10.7$) pairs.
In the {\it lower left} panel we see an obvious difference
between the colour of SAT-CEN pairs in SAT sample S1,
with a median difference of 0.03 for
massive (${\rm log(M_{star}/h^{-2}M_{\odot}}>10.7$) pairs and 0.3 for low mass
(${\rm log(M_{star}/h^{-2}M_{\odot})}<10.3$) pairs. 

To evaluate whether the above biases are statistically
significant or not, we match our CEN sample with mock samples created
from itself with random changes on ${\rm M_{star}}$ and colour according to 
the measurement uncertainties. The match is done with the same criteria 
used for constructing the SAT sample S2 and repeated 20 times to produce 
a distribution of medians of $\Delta {\rm M_{star}}$ and $\Delta ^{0.1}(g-r)$ 
per ${\rm M_{star}}$ bin between the CEN sample and mock samples.
We plot the results in Figure \ref{fig:matching} with circles 
and errorbars showing 
the mean and 3$\sigma$ deviation respectively.
We can see that the biases
of our real CEN-SAT matches are within
3$\sigma$ deviations of the self-matching results, except in the {\it lower
left} panel, where the bias at ${\rm log(M_{star}/h^{-2}M_{\odot})}<11$ are
much larger than the 3$\sigma$ deviations. This result suggests that
the ${\rm M_{star}}$ and colour distributions of our SAT samples are not
significantly different from the distributions of our CEN sample, except
for the case of colour in our SAT sample S1, where we find that 
SATs are redder than their CEN counterparts as \citet{vandenbosch08a} finds.  
However, the colour difference between
low mass CENs and SATs in sample S1 ($\Delta ^{0.1}(g-r) \sim 0.3$) is
larger than the average difference
of $0.1$ in the same stellar mass region found by \citet{vandenbosch08a}.
The reason for this disagreement
could be the small number statistics in our SAT sample.
We conclude that
the SATs from sample S2 is effectively matched in ${\rm M_{star}}$ and colour with our
CEN sample, while the SATs from sample S1 are only matched in ${\rm M_{star}}$ and
have a systematic redder colour than the CEN sample.

\section{Measuring the Structure of Galaxies}
\label{pipeline}

The primary aim of this paper is to quantify the structural properties of CENs. In 
particular, we measure the shape and size of the 2-D luminosity profile of each 
galaxy using GALFIT \citep{peng02}. This code fits a parametric model to the surface 
brightness profile of a galaxy image and outputs a set of best-fitting
parameters. For our 
analysis we adopt the \citet{sersic68} model to describe the 
surface brightness at radius $r$ of a galaxy, 
$\Sigma(r) = \Sigma_e e^{-b_n[(r/r_{50})^{1/n}-1]}$, where $r_{50}$ 
is the half-light radius of a galaxy, $\Sigma_e$ is the surface brightness 
at $r_{50}$, $n$ is the 
\sersic index, and $b_n$ is coupled to $n$ such that half of the total flux 
of a galaxy is within $r_{50}$ (for $0.5<n<10$, $b_n \approx 2n-0.327$). 
In addition to $r_{50}$ and n, GALFIT outputs the best-fitting total 
magnitude ${\rm m_{tot}}$, axis ratio $b/a$, and position angle $\theta$.
The \sersic profile is routinely used for galaxy 
structure analysis to provide the half-light radius and the 
\sersic index measures the galactic light profile shape; e.g., $n=4$ is the
de Vaucouleurs' 
$r^{1/4}$ profile and $n=1$ is the exponential disk profile.
We choose GALFIT as our fitting tool because it 
can simultaneously fit \sersic profiles to several galaxies in one image, which is
advantageous for galaxies in dense environments,
where galaxies have a high chance to 
be overlapping with one other. In what follows, we
outline our image fitting pipeline, test it with simulated galaxies,
discuss technical issues, and estimate the effects
that background sky estimation has on parameter uncertainties.

\subsection{The Galaxy Fitting Pipeline: Modified GALAPAGOS}
\label{galapagos}

To run GALFIT on each galaxy in our sample, we require a postage stamp 
image with an appropriate size to measure structure over the full extent of
an object,  
the point spread function (PSF), the initial guess for the fitting parameters,
and an estimate of the background sky level.
In our pipeline, we start from the fully-processed SDSS imaging frames,
which have a size of $2048\times1042$ pixels.
We employ GALAPAGOS (Barden in prep.) to process the whole image frame and 
to provide the needed information to GALFIT.  GALAPAGOS was originally designed 
to facilitate fitting large galaxy data sets based on HST/ACS images.
We have modified this routine to work on SDSS images.
In brief, GALAPAGOS takes the following steps: 
(1) it detects neighbouring sources and produces image masks; (2) it cuts out
postage stamps for
detected sources; (3) it prepares an input file for GALFIT; and (4) 
it estimates local sky values for target galaxies.
We outline the details of these steps below.

\textit{Source Detection and Initial Fit Parameter Guesses:} 
We use SExtractor \citep{bertin96} to detect and mask
nearby companion sources in the SDSS image of each galaxy that we desire to fit,
and to provide initial fit parameter guesses for the primary galaxy and any
close neighbours that will be simultaneously fit.
SExtractor provides useful estimates of 
galaxy properties (magnitude, size, axis ratio, position angle) for calculating
the initial guesses for the GALFIT parameters. 
A set of configuration parameters defines
how SExtractor detects sources. After tuning, we find that for SDSS
{\it r}-band images the 
following configuration works best: DETECT\_MINAREA=25, DETECT\_THRESH=3.0 and 
DEBLEND\_MINCONT=0.003. 
This configuration provides a good trade off between detecting
and deblending most bright and extended sources without artificially deblending
galaxies with strong substructures into multiple sources. 
Since our goal is to study bright galaxies in 
groups or clusters, we keep this configuration to allow a high success rate on
bright galaxies. To distinguish whether companions are stars or galaxies,
all sources with SExtractor flag STAR\_CLASS$<$0.9 are classified as galaxies. 
However, a small fraction of sources may be misclassified using this
criterion, and we will discuss the effect of this misclassification on our
fitting results in \S\ref{issues}.

\textit{Postage Stamp Images:} GALAPAGOS produces a rectangular postage stamp
centred on each galaxy of interest. 
The purpose of postage stamps is to reduce the CPU time for fitting one galaxy.
The postage stamps are cut to a size that is large enough to ensure that
the outer light profile
will be fit. GALAPAGOS uses parameters in the SExtractor catalogue to determine
the X and Y dimension in pixels of the postage stamps for each 
object:
\begin{equation}
X = 2.5 \times a \times r_{Kron} \times (|sin(\theta)| + (1-e)|cos(\theta)|),
\end{equation} 
\begin{equation}
Y = 2.5 \times a \times r_{Kron} \times (|cos(\theta)| + (1-e)|sin(\theta)|),
\end{equation} 
where $a \times r_{Kron}$ is the 
Kron radius along the major axis in units of pixels, $\theta$ is
the position angle, and $e$ is the ellipticity.

\textit{GALFIT Inputs:} For running GALFIT, GALAPAGOS produces an input file
of initial parameter guesses for the fitting parameters based on the SExtractor
output as follows: ${\rm m_{tot,i}=MAG\_BEST}$ for the apparent $r$-band 
magnitude, ${\rm r_{50,i}} = 0.162 r_{flux}^{1.87}$, where 
$r_{flux}$ is FLUX\_RADIUS; ${b/a}_i=1-e$, where $e$ is ELLIPTICITY; 
and 
$\theta_i={\rm THETA\_IMAGE}$. We start with an initial \sersic index of $n=1.5$. 
We note that for this analysis, we do not use the higher fitting modes, 
such as diskyness and boxiness,
offered with the GALFIT software.
Nearby companions within 1.5 times the SExtractor Kron aperture of the target 
galaxy are fit simultaneously with a \sersic model using initial parameters 
also determined as described above. Companions further away are masked out
using the masks provided by SExtractor. 
In addition to the input file, GALFIT requires a PSF image to convolve with
the model image. The PSF at the 
centre of the target galaxy is extracted from the SDSS photo pipeline
by employing a SDSS published tool 
readAtlasImages\footnote[1]{http://www.sdss.org/dr4/products/images/read\_psf.html}. 

\textit{Background Sky Estimation:} The background sky level is a critically
important ingredient in galaxy image fitting. For example, an overestimation of
the sky can result in flux, size, and \sersic index 
underestimation. GALAPAGOS includes a 
sophisticated way to measure the local sky around a galaxy, 
which is demonstrated 
to be successful for ACS images \citep{haussler07}. However, GALAPAGOS uses a 
hierarchical iteration for
fitting galaxies, from bright to faint, over the whole 
frame to isolate sky pixels. Rather than using this CPU-intensive approach,
we rely on the SDSS,
which provides useful and well-tested sky estimates. The SDSS global sky 
is considered to be a good measure of the background sky level for studying the 
structure of SDSS galaxies \citep{vonderlinden07}. 
We treat the background sky as a fixed flux pedestal 
during the fitting to reduce degeneracies between the sky and 
the outer isophotes of high-$n$ models.

\subsection{Testing the Pipeline with Simulations}
\label{pipetest}

We test our SDSS image fitting pipeline by running it on 850 simulated galaxies.
The goals of this test are two fold: (1) to estimate random and bias
of the structural parameters returned by our fitting pipeline;
and (2) to confirm that using 
the SDSS global sky does not produce bias. We compare
the actual properties that define each simulated galaxy
(input) to the corresponding fit result (output) following 
Figure 9 in \citet{blanton05nyu}. 

For our tests,
we use the SDSS-based simulations of \citet{blanton05nyu}. Briefly, each 
simulated galaxy is an 
axisymmetric \sersic model. The simulation sample has a range of input 
parameters matching the NYU-VAGC Large-Scale Structure (LSS) sample
from SDSS DR4.
Each simulated galaxy is converted to raw data units, scaled to the SDSS pixel 
size, and convolved with the PSF at its position. After adding Poisson noise,
the resulting image is placed into an actual SDSS image at a random location.
Generally, about 60 simulated 
galaxies were added to each $2048\times1024$ pixel image.

\begin{figure*}
\center{\includegraphics[scale=0.8, angle=0]{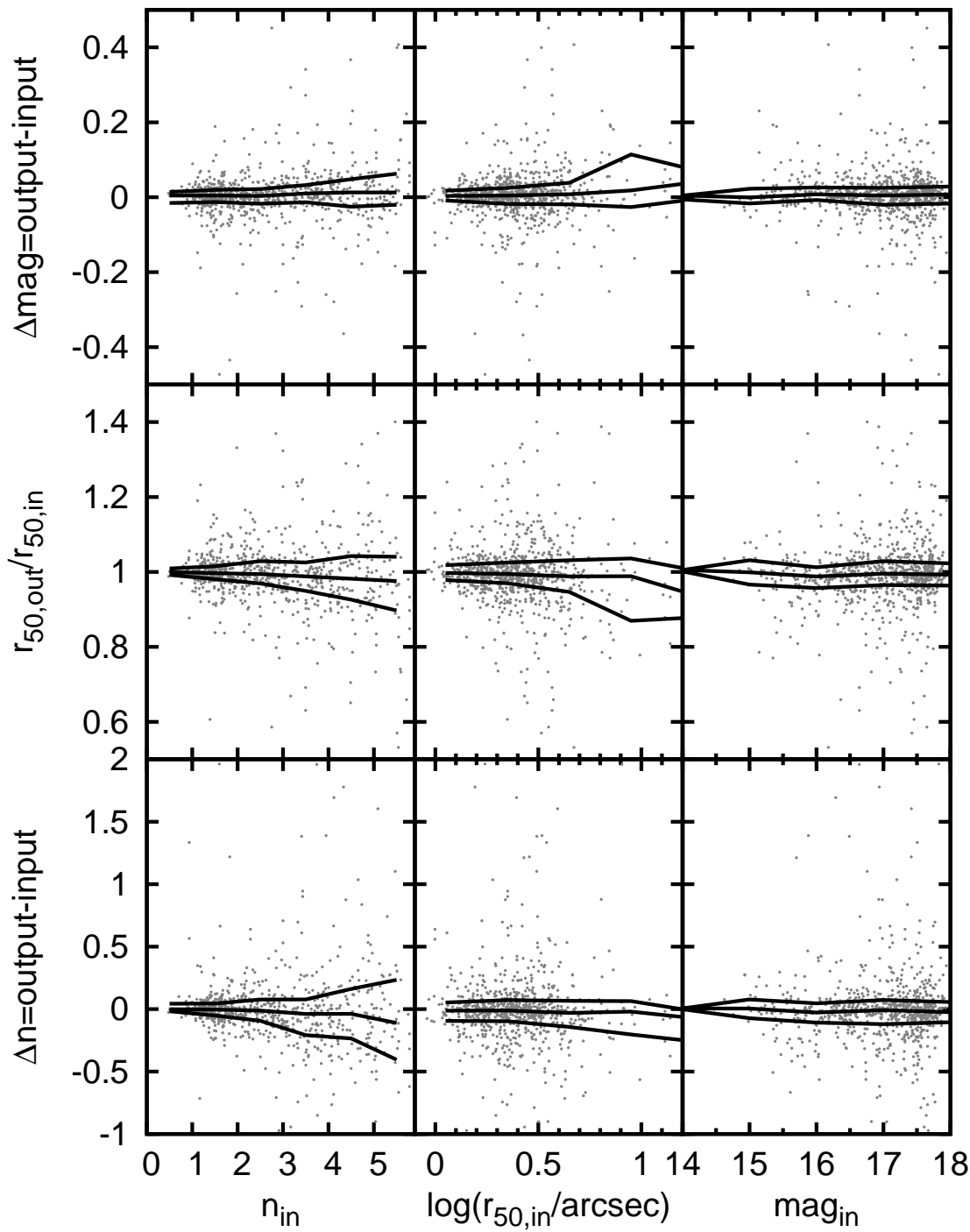}}
\caption[]{A comparison between input and output parameters when using GALFIT to
fit \sersic models to 850 simulated \sersic galaxies. We
plot the difference (output-input) as a function of input for
three important \sersic parameters: the index $n$, the half-light
radius $r_{50}$, and the total magnitude. In each panel, grey points represent
individual simulated galaxies. The solid
lines show the 3rd quartile, median and 1st quartile (from top to bottom). 
\label{fig:simmatrix}}
\vspace{-0.2cm}
\end{figure*}

The results of applying our pipeline to the SDSS simulations are shown
in Figure \ref{fig:simmatrix}. 
We plot the output-input offset as a function of input for three important
parameters: the total magnitude ${\rm m_{tot}}$, the 
half-light radius $r_{50}$, and the
\sersic index $n$. The results demonstrate that our pipeline 
successfully recovers these structural 
parameters of simulated \sersic galaxies with almost no offset and only a
small scatter.  For half of the simulation sample
the pipeline returns structural measurements within
10 percent of the true value. 
As expected, there is a 
larger parameter offset scatter for simulated galaxies with 
$n_{in}>4$ than for those with $n_{in}<4$ (see \citet{haussler07}). 
We see a slightly
increased scatter with magnitude owing to the lower S/N of fainter galaxies. 
We note that the good agreement between input and output parameters also
demonstrates that the SDSS global sky is a good choice for galaxy
image fitting. We address the effects of sky uncertainties on our structural properties
measurements in more detail in \S\ref{covar}. To the degree that \sersic profiles provide a reasonable fit to the true
light distribution of galaxies, the good performance of our pipeline on simulated galaxies
is promising for the analysis of real galaxies.

\subsection{\sersic Fitting of the CEN Sample}
\label{issues}

We apply our pipeline to fit each galaxy in the CEN sample with a \sersic 
model.  To evaluate the fit quality we visually 
checked a random subsample of 200 CENs and find that about 15 percent of our
fits suffer from a variety of technical issues. These issues include 
stars misclassified as galaxies and unreliable fits to companion sources.
Here we discuss what portion of our
fitting results are affected by these issues and 
how we can correct for them.

The accuracy of star/galaxy classification affects the fitting quality. 
Although 
CEN galaxies can always be correctly classified, a small portion of their 
companions may suffer from misclassification.
Some galaxies are misclassified as stars and some stars are misclassified as 
galaxies, but only the latter has a severe impact on our fitting.
Fitting a \sersic model to a stellar profile typically results in a very large
$n$ with an overestimated extent
and an overestimated flux. As a result, this false extension takes
light away from the whole image and 
hence causes an underestimate of ${\rm m_{tot}}$, $r_{50}$
and $n$ for the target galaxy.  Conversely,
companion galaxies misidentified as stars
and fit by a PSF are spatially compact and
faint, thus their improper treatment has
little impact on the measured structure or flux of the target galaxy.
We estimate that less than 5 percent of our CEN sample (in a random sense) suffer
from the issue of companion stars being misclassified as galaxies.

A drawback to fitting several sources simultaneously is that an unrealistic
model fit 
for a companion will spoil the fitting quality of the target galaxy.
We develop an empirical set of criteria for identifying bad companions such 
that any companion being fitted with 
$\sigma_{{\rm a_{50}}}/{\rm a_{50}}>0.3$ and $n>8.0$ is identified as a 
bad companion,  where ${\rm a_{50}}$ is the semi-major axis 
${\rm a_{50}}=r_{50}/\sqrt{b/a}$, $b/a$ is the axis ratio, and
$\sigma_{\rm a_{50}}$ is the internal GALFIT
error of ${\rm a_{50}}$. 
Such
bad companions are usually small and/or low surface brightness galaxies 
and tend to be 
fit with an overestimated size, \sersic index and magnitude, resulting
in a severe underestimate of the same quantities for the target galaxies. 
We examine SExtractor magnitude differences between bad companions ($m_{BC}$)
and target galaxies ($m_T$) and find
that bad companions, as defined here, 
are mostly fainter than their corresponding targets. Actually, the majority 
of bad companions have $\Delta m=m_{BC}-m_T>2.5$ and the distribution 
of $\Delta m$ is peaked at $\Delta m \sim 3.5$.
Thus we simply mask out all $\Delta m>2.5$ bad companions and refit 
the image. We iterate this procedure until all the targets have no more bad 
companions. About 8 percent of our CEN sample start with bad companions and we
correct all of them as described above.
Some of the target CENs have $\Delta m<2.5$
bad companions within a few arcseconds and any masking might adversely 
affect the fit. Hence instead
we opt to exclude those target galaxies from our CEN sample. The 
fraction of CENs having such bad companions depends on halo mass: from 7 percent
at high-masses to 3 percent at low-masses.
In total, 32 CENs are excluded from our
sample owing to problems with bad companions.

\subsection{Sky Uncertainty and Parameter Covariance}
\label{covar}

Although the SDSS global sky works as a good choice of sky background for
fitting isolated galaxies, as the simulations show,
the real sky is difficult to measure especially for CENs in 
dense environments. In this subsection, we estimate the uncertainty in the
SDSS global sky values and evaluate how 
this uncertainty translates into fit parameter errors.

We use the difference 
between the SDSS global sky and mean sky to characterise the uncertainty in
the sky measurement.  SDSS measures the global sky as the median data counts
(ADUs) from every pixel of the source-subtracted
frame after sigma-clipping. Besides the global sky, SDSS also provides a 
mean sky for each frame and a local sky for each detected object in the frame.
SDSS measures the background in sub-frame boxes of
of $256 \times 256$~ pixels centred every 128 pixels. 
The local sky of each object
in a SDSS frame is an interpolation of the sub-frame background values
at the position of the object centre.
The mean of all sub-frames is the mean 
sky for a frame. 

\begin{figure*}
\center{\includegraphics[scale=0.8, angle=0]{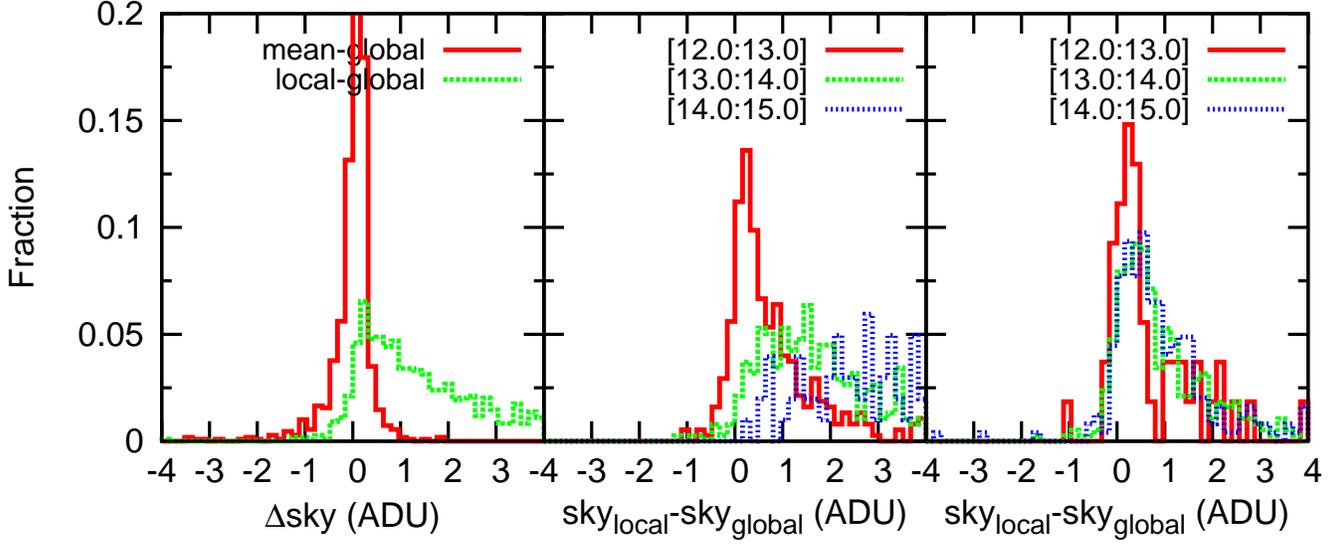}}
\caption[]{Distributions of differences between various sky measurements.
\textit{Left}: The red line shows the difference between the SDSS mean and 
global sky values for our CEN sample and the green line shows the difference
between the SDSS local and global measurements. \textit{Middle}: The 
local-global difference for CENs divided into halo mass bins. \textit{Right}:
Same as the middle panel, but for SATs.
\label{fig:dsky}}
\vspace{-0.2cm}
\end{figure*}

In the {\it left} panel of 
Figure \ref{fig:dsky}, we plot the
difference between the SDSS global sky and mean sky for
our CEN sample. The mean sky is somewhat larger than the global sky on average 
because individual 
sub-frames may suffer contamination from large and bright sources. 
We also plot the difference between the SDSS global and local sky measurements
for our sample in the left panel. These
differences are much larger than that between the global and mean 
sky because the local sky is heavily contaminated by large and bright sources. 
Furthermore, in the {\it middle} panel of Figure \ref{fig:dsky},
we show that the overestimates of the 
local sky are halo-mass dependent for CENs.
The local sky at the centre of massive haloes 
tends to be higher than that in less massive haloes. Since the real sky 
background should be independent of the properties of haloes, this dependence
implies 
that local sky measures for CENs suffer from increased contamination from 
the higher density of galaxies in larger haloes. 
In contrast, SATs display little dependence on halo mass, even
though their local sky is also overestimated,
as shown in the {\it right} panel of Figure \ref{fig:dsky}.
We attribute this lack of halo dependence to the fact that SATs 
are found over a range of halo-centric positions and hence at various 
galaxy densities.  It is also possible that intra cluster light (ICL)
at the centre of massive haloes can contribute to the difference
between CEN and SAT sky estimates as we show in Figure \ref{fig:dsky}. 
Given the large overestimates 
and the halo dependence of the SDSS local sky, we will not use it as
a measurement of the sky background.
Moreover, we conclude that the SDSS mean sky not only provides a robust 
alternative measurement of the sky background 
but it also tells us in which direction the sky measurement might be biased. 

\begin{figure}
\center{\includegraphics[scale=0.6, angle=0]{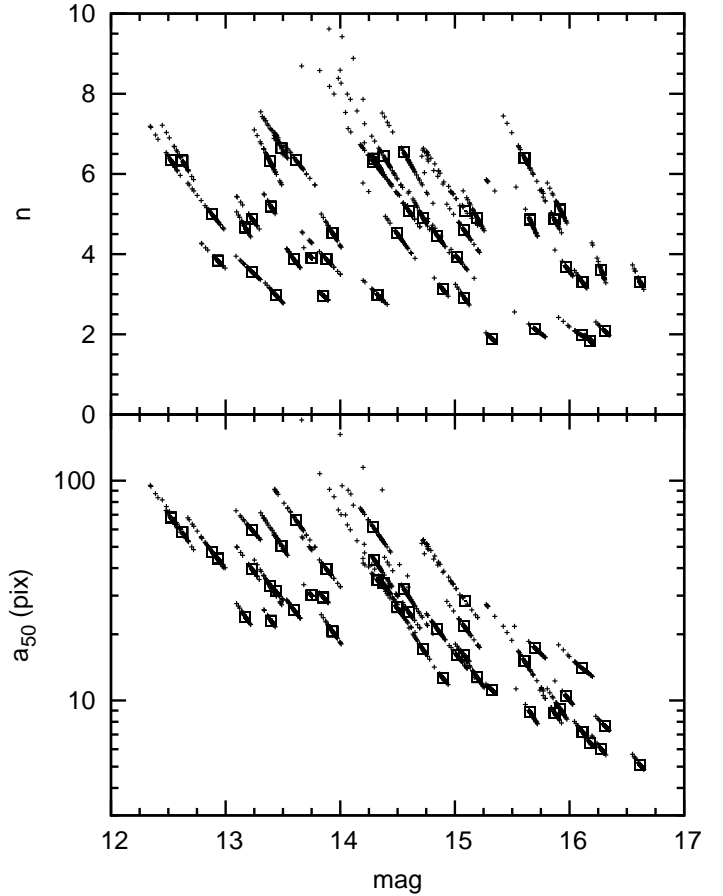}}
\caption[]{The covariances among best-fitting total magnitude, \sersic index and 
${\rm a_{50}}$ as a result of sky uncertainty for 45 representative CENs. Boxes
represent the GALFIT result using the SDSS global sky. For each galaxy, the 
series of (50) crosses represent fits using sky background levels taken
from Monte Carlo samplings of the sky as described in the text.
\label{fig:mcmc}}
\vspace{-0.2cm}
\end{figure}

To study the effect of sky uncertainties on the \sersic fitting of
actual galaxies, we first
select a representative subset of 45 CENs from our sample that span
a $3\times3$ matrix in ${\rm M_{halo}}$-$n$ space as follows:
\begin{itemize}
 \item for ${\rm log(M_{halo}/h^{-1}M_{\odot})}$=[12.0:12.5], {\it n}=[1.8:2.2], [3.3:3.7], [4.8:5.2] 
 \item for ${\rm log(M_{halo}/h^{-1}M_{\odot})}$=[13.0:13.5], {\it n}=[2.8:3.2], [4.3:4.7], [6.3:6.7] 
 \item for ${\rm log(M_{halo}/h^{-1}M_{\odot})}$=[14.0:14.5], {\it n}=[3.8:4.2], [4.8:5.2], [6.3:6.7] 
\end{itemize}
Next, we randomly select five galaxies from each bin of the above matrix.
For each galaxy, we Monte-Carlo sample 50 values from the distribution of 
the mean-global difference shown by the red line in the {\it left} panel of 
Figure \ref{fig:dsky}. These values are added to the global sky and the 
galaxy is 
refit using these new background levels. This procedure provides a distribution 
of fitting parameters caused by the uncertainty in measuring the sky. For this 
subset of 45 CENs, we plot distributions around the mean value of three
important fit parameters ($m_{\rm tot}$, $n$ and ${\rm a_{50}}$)
in Figure \ref{fig:mcmc}.

The distributions in
Figure \ref{fig:mcmc} clearly demonstrate a covariance between best-fitting
parameter values and the choice of sky.
The boxes represent our original results using the SDSS global sky. The 
series of (50) crosses, which form a short and nearly straight line through 
each box, represent the fitting results using the Monte Carlo
sampling of the sky background as described above. The correlations between
${\rm m_{tot}}$ and ${\rm a_{50}}$ (bottom) and ${\rm m_{tot}}$ and $n$ (top)
show the strong degeneracy of these parameters in the \sersic model.
If we increase the sky background,
the best-fitting flux decreases to keep the total (galaxy + sky) flux constant, 
while 
the best-fitting $n$ and ${\rm a_{50}}$ decrease. 
Conversely, decreasing the 
sky background level results in larger best-fitting values for
${\rm m_{tot}}$, $n$, and ${\rm a_{50}}$. 
We find that the covariance is more severe for galaxies with higher $n$.
This effect shows that the sky estimate is very crucial 
for producing accurate \sersic profile fits.  This is especially true
for galaxies with $n>4$ because the flat, extended wings of the profile are
sensitive to sky uncertainties. The strength and direction of these covariances
must be accounted for when analysing the size-luminosity
relation, as we will discuss in \S\ref{size}

\section{The Structure of CENs}
\label{results}

In this section we explore the
{\it r}-band structural properties of our CEN sample.
CENs are the most massive members of the SDSS groups and clusters,
and are presumed to be located at the dynamical centre of the host halo. 
We focus on the structural shape (characterised by the \sersic index $n$) and the 
size (characterised by the half-light radius $r_{50}$) of the \sersic profile 
of CENs. 
We study the relationships between these parameters and galaxy stellar 
masses and host halo masses to investigate which factor is more related to 
the structure of CENs. We also compare the structural parameters of
CENs and SATs using our two matched SAT samples (see \S\ref{sample} for details)
and
study whether the central halo location impacts the shape or size of a galaxy.

To evaluate the accuracy of our measurements of galaxy structural parameters
and total flux using our GALFIT pipeline, we compare our fit results 
with the
one-dimensional (1D) \sersic fitting parameters based on the NYU-VAGC analysis
\citep[][hereafter 'NYU-VAGC']{blanton05nyu}.
The NYU-VAGC
\sersic parameters have been widely used in the study of galaxy morphology
and size \citep[e.g.][]{shen03,blanton05a,maller08}. We find that
for high \sersic index (our GALFIT $n>3.0$) galaxies, NYU-VAGC underestimates 
the \sersic index by about 1.3 and underestimates the total magnitude 
by about 0.4 mag compared to our results. There are two reasons for these
underestimates: (1) NYU-VAGC's 1D 
profile fitting systematically underestimates the \sersic index and total
magnitude for high $n$ galaxies, as shown by Figure 9 of \citet{blanton05nyu};
(2) NYU-VAGC uses the SDSS local sky, which overestimates the background sky level
and hence results in underestimates of $n$ and the total magnitude. Furthermore,
for low \sersic index (our GALFIT $n<3.0$) galaxies,
NYU-VAGC overestimates the \sersic 
index but still underestimates the total magnitude by about 0.2 mag.
Besides using the local sky, NYU-VAGC's azimuthally-averaged 1D fitting 
procedure systematically overestimates the \sersic index for highly inclined 
galaxies, which are mostly disks. We also compare both our GALFIT results and 
the NYU-VAGC results with the Petrosian quantities
from the SDSS photometric pipeline.
By following the formalism of \citet{graham05}, we find that under our 
working assumptions (a \sersic model and a global sky), our GALFIT fitting 
pipeline more accurately measures the structural properties of SDSS galaxies
then NYU-VAGC's. The details of this comparison are in Appendix \ref{petro}.

\subsection{The \sersic Index of CENs}
\label{sersic}

With reliable and accurate measurements of galaxy structural properties
from our fitting pipeline in hand,
we turn to the study of how the \sersic index of CENs is related to
their stellar mass and their environment. 
The \sersic index is widely used to characterise the 
profile and concentration of 
galaxies in both observational \citep[e.g.,][]{graham96,blanton05a,macarthur03,dejong04,allen06} and
numerical \citep[e.g.,][]{naab06c,aceves06} studies.
Some argue that the 
morphology-density relation implies that the structure of galaxies is affected by 
the environment. Others argue that the structure of galaxies 
depends strongly on stellar mass but only weakly on the environment 
\citep[e.g.,][]{hogg04,kauffmann04,vanderwel08}.
In this section, we explore the dependence of the 
\sersic-index distribution of CEN galaxies
on both \mstar and ${\rm M_{halo}}$.

\subsubsection{Dependence on Stellar Mass}
\label{mstardep}

\begin{figure*}
\center{\includegraphics[scale=0.85, angle=0]{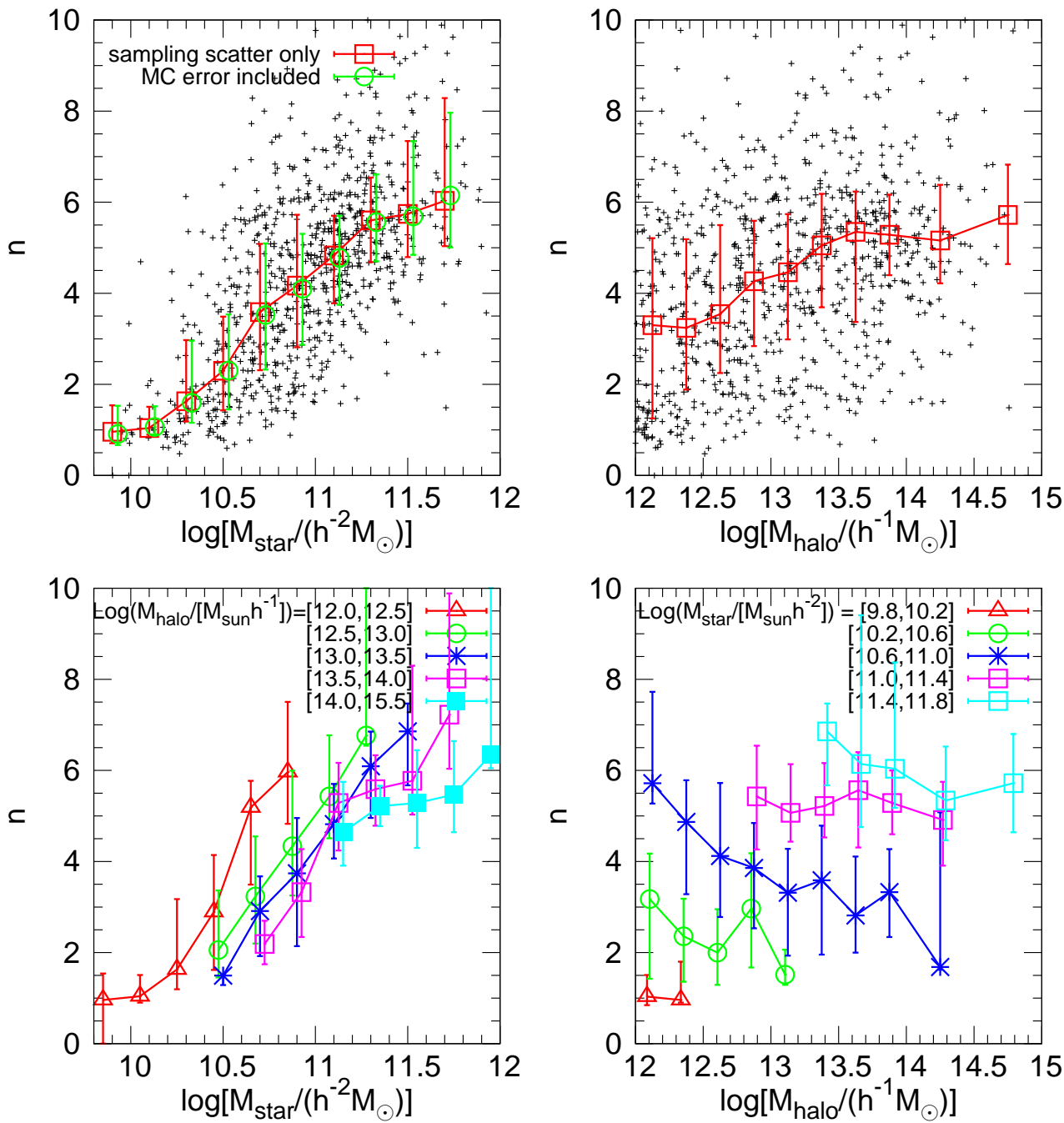}}
\caption[]{\textit{Upper left}: the relation between \sersic index and stellar mass for
our CEN sample. Black crosses represent the best-fitting $n$ for 
individual CEN galaxies. The red line
connects the
median (open squares) of the $n$ distribution in each \mstar bin. The green
line and open circles show the median once the $n$ and $m_{\rm tot}$ fitting
uncertainties owing to sky are folded into the distribution 
(see text for details).  The errorbars show the first and third quartiles.
\textit{Upper right}: the relation between the best-fitting \sersic index and halo mass for 
our CEN sample. Black crosses and red lines with errorbars are done as in
the left panel.
\textit{Lower left}: the $n$-\mstar
relation split into different bins of fixed halo mass as indicated by key.
\textit{Lower right}:
the $n$-\mhalo relation split into different bins of fixed stellar mass as
indicated by the key.
In both lower panels, the symbols and errorbars provide the median, 1st, and
3rd quartiles.
In the {\it lower} panels, only the best-fitting results are used. 
\label{fig:nvsms}}
\vspace{-0.2cm}
\end{figure*}

In the {\it upper left} panel of Figure \ref{fig:nvsms}, we plot the
\sersic index of CENs as a function of
their stellar mass. The \mstar values for our CEN sample are
calculated following the formula from \citet{bell03b}:
\small
\begin{equation}
{\rm log}[{\rm M}_{\rm star}/{\rm M}_{\sun}]= - 0.306 + 1.097^{0.0}(g-r) - 0.15 - 0.4(^{0.0}M_r-4.67) ,
\label{eq:mstar}
\end{equation}
\normalsize
where the constant 0.15 corrects to a \citet{kroupa01} IMF, and
$^{0.0}(g-r)$ is the Petrosian colour from the NYU-VAGC
shifted to the $z=0$ rest-frame using the \citet{blanton03b} $K$-corrections and
correcting for Milky Way extinction using the \citet{schlegel98} dust maps.
The absolute $r$-band magnitude
$^{0.0}M_r$ is extinction and K-corrected in the same manner, but we use
the total flux from our fitting for this quantity rather than
the Petrosian magnitude in the SDSS pipeline.
We show in Appendix \ref{petro} that our fitting procedure better
recovers the total magnitude of galaxies than the Petrosian photometry. 
The red symbols with errorbars in Figure \ref{fig:nvsms} show the 
median and the 1st and 3rd quartile of the $n$ distribution in \mstar
bins with a 0.25 dex width. From the plot we see
that the median \sersic index is a strong function of stellar mass:
low-mass CENs have low $n$ (disk-like) and high-mass
CENs have high $n$ (spheroid-like). Yet, there is large scatter in the
relation between $n$ and \mstar, especially for CENs with 
$3\times 10^{10} < {\rm M_{star}/h^{-2}M_{\odot}} < 10^{11}$, where the \sersic
index ranges from $1<n<6$ or larger. Higher-mass CENs tend to have $n>3$.
 
We also show how the uncertainty and covariance of the GALFIT $n$ values
affect the $n$-\mstar relation in the
{\it upper left} panel of Figure \ref{fig:nvsms} 
(green symbols).  The red symbols in the plot only take into
account of the scatter from our fits. However, as we showed in 
\S\ref{covar}, 
there is a strong covariance between the fit parameters $n$ and $m_{\rm tot}$
and the sky uncertainty, and we use $m_{\rm tot}$ to estimate stellar mass.
To evaluate how this covariance may change the above $n$-\mstar relation, 
we re-calculate the relation using the results from the
representative subset of 45 CENs in \S\ref{covar}.
For each CEN in our sample we randomly select one CEN from the nearest
five in $n$ space among the representative subset, and we assume that each
CEN has the same $n$ and $m_{\rm tot}$ covariance as its matched companion.
In this way, we construct a probability distribution of $n$ and \mstar
(from $m_{\rm tot}$) for all 911 CENs. The median, 1st and
3rd quartiles of this distribution are shown by the green symbols and 
errorbars in the {\it upper left} panel of Figure \ref{fig:nvsms}.
The new relation shows little difference compared
with the original because the parameter uncertainties owing to sky
are smaller than
the intrinsic scatter in our sample (red symbols). For example,
the average relative scatter between the 1st and 3rd quartiles of $n$ 
owing to the background sky level uncertainty
is only $\Delta n/n \sim 0.2$, while the measured scatter is
$\Delta n/n \sim 0.75$ and hence dominates.

\subsubsection{The Dependence on Host Halo Mass}
\label{mhalodep}

In the {\it upper right} panel of Figure \ref{fig:nvsms},
we show the dependence of 
the \sersic index of CENs on their host halo mass. Recall that
\mhalo is calculated by 
matching the rank of the total stellar mass of groups and clusters with that of 
dark matter haloes from numerical simulations (see Y07 for details). We
find that $n$ depends only weakly on \mhalo and that the scatter of the
$n$-\mhalo relation is large.
For CENs in haloes with ${\rm log[M_{halo}/h^{-1}(M_{\odot})]<12.5}$, 
we find a median value of $n=3$ and a relative scatter of
$\Delta n/n \sim 1.3$. For
massive haloes with ${\rm log[M_{halo}/h^{-1}(M_{\odot})]>14.0}$ 
the median is $n=5$ and 
the relative scatter is $\Delta n/n \sim 0.6$.

The $n$-\mhalo relation, although weak, suggests that the halo mass may also affect the structure of 
CENs. However, from the right panel of Figure \ref{fig:samplecheck} we know that
the stellar mass of CENs also depends on the halo mass in 
the sense that CENs in massive haloes tend to have larger stellar masses
than CENs in smaller haloes. Given this dependence and the strength
of the $n$-\mstar relation, it is tempting to rule out any dependence of
$n$ on halo mass. To address this,
we attempt to remove any \mstar-\mhalo dependence from 
both the $n$-\mstar and $n$-\mhalo relations.

First, in the lower left panel of Figure \ref{fig:nvsms}, we plot
the $n$-\mstar relations for CENs in five halo mass bins, each 0.5 dex in width.
All the bins with less than six CENs are excluded to get better statistics.
The roughly \mhalo-independent relations all have a
slope and amplitude that is similar to the single $n$-\mstar relation for the 
full sample (red symbols in the {\it upper left} panel).
There is some evidence that the relations for different \mhalo bins
are somewhat offset along the \mstar direction in the
sense that CENs in less massive haloes tend to have larger $n$ than their 
counterparts in more massive haloes. 
But since the scatter in the relations are large,
we cannot draw any firm conclusions.
 
Similarly, we study the $n$-\mhalo relations of CENs in five different 
\mstar bins, each with a width of 0.4 dex, as
shown in the {\it lower right} panel of Figure \ref{fig:nvsms}.
We find that the $n$-\mhalo relations for each bin of fixed galaxy mass 
is different from the single $n$-\mhalo relation for the full sample.
For the most part, the relations are all flat, i.e. each CEN of a given
\mstar has a constant $n$, within the scatter,
independent of its host halo mass. We note that several of the 
fixed stellar mass relations have a small negative slope when considering
only the median values.
This slight trend is a manifestation of the small offsets among the
$n$-\mstar relations in
different \mhalo bins in the {\it lower left} panel. A much larger sample
is required to validate whether the profile shape of CENs, as measured
with \sersic fitting, has a small second-order dependence on halo mass.

\subsubsection{Discussion}
\label{bcgdiscuss}

We study the distribution of $n$ for our CEN sample and its dependence on \mstar and 
\mhalo and find that the \sersic profile shape
of CENs is strongly correlated with \mstar but only weakly
(or not at all) correlated with ${\rm M_{halo}}$. Low-mass CENs 
tend to have shallower, disk-like profiles, while 
massive CENs have steeper, more spheroid-like shapes. 
This $n$-\mstar relation holds for CENs from different
\mhalo bins with almost the same slope and amplitude. On the other hand,
CENs have $n$ values that depend very weakly, if at all, on global
environment as defined by 
${\rm M_{halo}}$, which is consistent with other observations
\citep{kauffmann04,vanderwel08}.
This correlation disappears (or even 
becomes an anti-correlation) if we divide 
our CEN sample into different \mstar bins.
This suggests that \mhalo has almost no 
effect on the shape or concentration of CENs. The key factor  that determines
$n$ of a CEN is its stellar mass.
Given the relationship  between the masses of CENs and their host halo masses,
the phenomenon of CENs in massive haloes having a more concentrated structure
(as implied
by morphology-density relation) can be explained simply by (1) an intrinsic
$n$-\mstar correlation, and (2) massive 
CENs living preferentially in massive haloes (i.e., mass segregation).

We find low-$n$ (disk-like) CENs in haloes spanning a large range in
halo mass
(from $10^{12}$ to ${\rm 10^{14}h^{-1}M_{\odot}}$), 
which rules out a distinct (special) \mhalo 
for producing spheroids. Likewise, finding high-$n$ (early-type) CENs over the
same range of environments suggests that major mergers, or some
other processes that transforms disk galaxies into spheroid-like
galaxies, occur at the centres of haloes with a wide range of masses.
Finally, it is unclear whether the halo plays a small,
secondary role in determining the overall profile shape of its CEN as hinted at
in the bottom panels of Figure \ref{fig:nvsms}.
The \mstar-\mhalo relation in Figure 1 shows that CENs
of a fixed stellar mass reside in haloes with a range of masses.
We speculate that
the tendency for CENs of a fixed stellar mass to be more concentrated
in smaller haloes and less so in
larger haloes may tell us something about the recent accretion history
of the host and its impact on the stellar mass growth of the CEN.
Halos grow through the accretion of, and the merging with, other haloes.
Briefly, consider two haloes (A and B) that were equal in mass and 
contained similar central disk galaxies in the recent past. Since that 
time both haloes have doubled in mass, halo A by a single major merger 
with a comparable halo, and B by accreting many minor subhaloes. Owing 
to dynamical friction timescales, the disk CEN in halo A will both (i) 
grow in mass faster than halo B's CEN and (ii) be transformed into a 
spheroid because it is doomed to experience a major merger with the 
CEN of the merging halo. Conversely, the mass of the CEN in halo B 
will take much longer to grow by the occasional minor merger with 
small infalling companions, and this process is much less likely to 
destroy its disk morphology. Therefore, when comparing haloes of the 
same mass, galaxies that grew through major mergers will be more massive 
and more concentrated. Likewise, when comparing CENs of the same stellar 
mass, those which grew through major mergers (i.e., more concentrated) 
will reside in lower mass haloes on average more than their disk-dominated 
counterparts as we see in the lower left panel of Figure \ref{fig:nvsms}.
Numerical modelling and a
more-detailed analysis on a much larger data set
could test these predictions.

\subsection{The Size of CENs}
\label{size}

\begin{figure}
\center{\includegraphics[scale=0.7, angle=0]{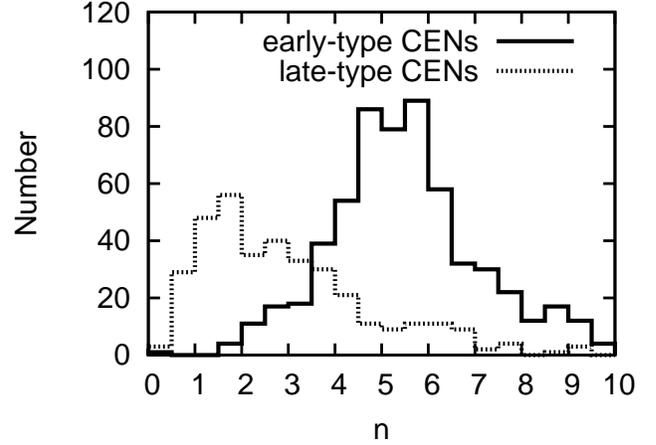}}
\caption{The distribution of \sersic index for our visual inspected 
early- (solid line) and late-type (dotted line) CENs. The \sersic 
index comes from our best fits. 
\label{fig:nsplit}}
\end{figure}

In addition to the \sersic index, the half-light size is an 
important characteristic 
of galaxy structure. It is well-established that galaxies follow
well-defined
size-luminosity ($r_{50}$-$L$) and size-stellar mass ($r_{50}$-${\rm M_r}$) relations 
\citep[e.g.][]{shen03,bernardi07,dutton07}, which are commonly used to
constrain galaxy formation and evolution theories.
In this subsection, we study 
the size-luminosity relation and the size-stellar mass relations of our CEN
sample and compare our results with those obtained by others. 
Our CEN sample contains both early and late-type galaxies.  Thus,
we visually inspect each galaxy and divide the sample into two types based
on whether spiral disk structure is present (late-type) or not 
(early-type). 
In Figure \ref{fig:nsplit} we plot the distribution of 
our best-fitting \sersic index for our visual inspected 
early- (solid line) and late-type (dotted line) CENs. Unlike employing a 
sharp cut at $n=2.5$ as in many studies, the majority of our early-type CENs
have $n>3.5$, while late-type CENs have $n<3.5$.
In what follows, we will discuss the early and late-type
size-luminosity and size-stellar mass relations separately.

\begin{figure*}
\center{\includegraphics[scale=0.85, angle=0]{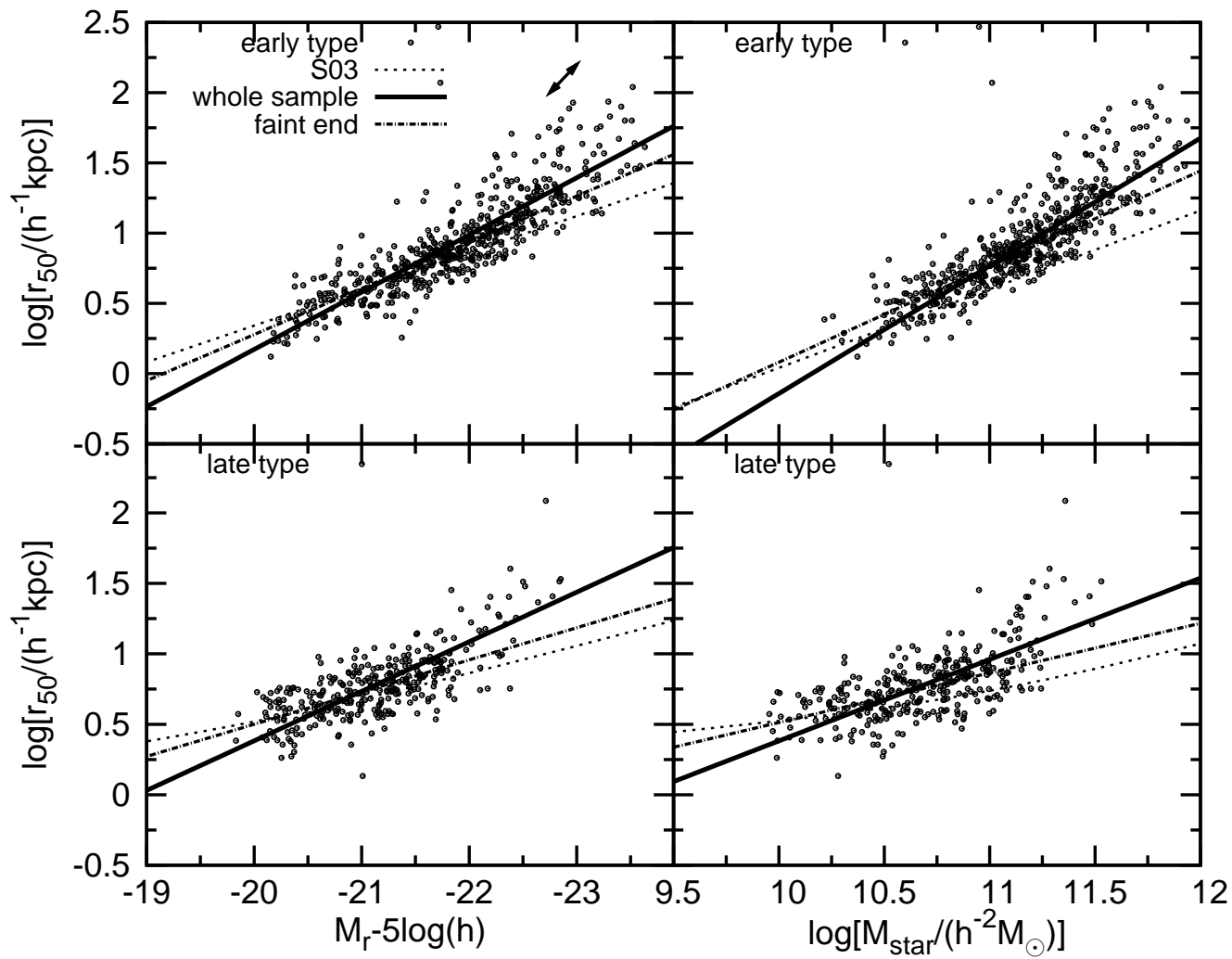}}
\caption{The size-luminosity (\textit{left}) and size-stellar mass 
(\textit{right}) relations for early-type (\textit{upper}) and 
late-type (\textit{lower}) galaxies in our CEN sample. The thick 
solid line in each panel is the best fit linear relation for
each morphological type over the whole luminosity/stellar mass range. 
The dashed line in each panel shows the linear relation by fitting 
the faint end (${\rm M_r-5log(h) \geq -22}$). 
For comparison, the
thin dotted line in each panel is the relation from 
S03 (see the text for details).
The arrow in the upper right panel shows 
the direction and amplitude of the average covariance between
${\rm M_r}$ and $r_{50}$ for bright (${\rm M_r-5log(h)<-22}$) CENs owing to background sky 
uncertainties (see \S\ref{covar} for details).
\label{fig:sizebcg}}
\end{figure*}

In Figure \ref{fig:sizebcg}, we show the size-luminosity relation of
early- (the \textit{upper left} panel) and
late-type (the \textit{lower left} panel) CEN galaxies. 
Here we use the half-light radius 
$r_{50}$ (in units of ${\rm h}^{-1}$ kpc) to represent the size of a galaxy. 
We have $r_{50} = \frac{\pi}{180} \times \frac{{\rm a_{50}}}{3600} \times
\sqrt{b/a} \times dA(z)$, where 
${\rm a_{50}}$ and $b/a$ is the semi-major axis and the axis ratio of 
our best-fitting model of the galaxy, 
$dA(z)$ is the angular diameter distance at redshift $z$ of the galaxy.
We calculate the absolute magnitude using the total magnitude from
our \sersic fit,
K+E corrected to $z=0.1$.
Following \citet[][hereafter S03]{shen03}, who also measured the $r_{50}$-$L$ 
or $r_{50}$-${\rm M_r}$ relation for SDSS galaxies, we
use linear regression to fit the $r_{50}$-$L$ relation, i.e.
${\rm log[{\it r}_{50}/(h^{-1}kpc)] = -0.4 \alpha M_{r} + \beta}$. We find
slopes of $\alpha=1.02\pm0.03$ for early-type CENs
and $\alpha=0.88\pm0.04$ 
for late-type CENs. The linear fits are shown by thick solid lines in
each panel.

At the bright end, ${\rm M_r-5\log{h}}<-22$, the relation steepens.
The origin of this steepening is unclear. It could be caused by 
the covariance between the semi-major axis ${\rm a_{50}}$ and 
the total magnitude ${\rm m_{tot}}$,
as shown in the upper panel of Figure \ref{fig:mcmc}.
Our tests in \S\ref{covar} show
that there is a very strong covariance between ${\rm a_{50}}$ and
${\rm m_{tot}}$, owing to uncertainties in the measurement of background sky
level, and that
this covariance becomes stronger for bright and large galaxies.
The covariance between ${\rm a_{50}}$ and ${\rm m_{tot}}$ will produce a
covariance between $r_{50}$ and ${\rm M_r}$ in Figure \ref{fig:sizebcg} since
${\rm M_r}$ is calculated from ${\rm m_{tot}}$.
The direction of the average covariance 
at the bright magnitude end, as shown by the arrow in the {\it upper left}
panel of Figure \ref{fig:sizebcg},
is almost parallel to the slope of the relation at the bright end.
Moreover, the amplitude of the covariance increases for brighter and 
larger galaxies as seen in Figure \ref{fig:mcmc}. 
Hence the covariance between size and total magnitude in the profile
fitting could contribute to the slope steepening at the bright end.

To account for any bias at the bright end, we refit
excluding sources with ${\rm M_r-5\log{h}<-22}$ and find
slopes of  
$\alpha=0.82\pm0.06$ (early-type) and $\alpha=0.57\pm0.04$ (late-type).
For reference,
a fit to the $r_{50}$-$L$ relation of early-type CENs brighter than -22 has
a slope of $\alpha=1.32\pm0.08$. We don't fit the bright end 
of late-type CENs, because there are only 
several galaxies.
Using the fit over the faint end (${\rm M_r-5\log{h} \geq -22}$), 
we find an
early-type $r_{50}$-$L$ relation with a much steeper slope 
than the $\alpha=0.65$ of S03 (the thin dotted line 
in the {\it upper left} panel of Figure \ref{fig:sizebcg}).
Our fit to the $r_{50}$-$L$ relation for late-type CENs fainter than -22 
is also qualitatively steeper than that of S03 (comparing 
the dashed and dotted line in the {\it lower left} panel), who 
fit the $r_{50}$-$L$ relation with a four-parameter model.
We note that the definition of early/late
types in S03 is different from ours.
They classified galaxies with a \sersic index 
$n>2.5$ as early-type and the others as late-type. 
The different definitions of early/late-type between S03 
and us could contribute to the measured slope differences.
We also try the same early/late-type classification as S03 
by using \sersic index from our best fits and refit the $r_{50}$-$L$ relation
for the faint end CENs. We find $\alpha=0.78\pm0.05$ for the slope of
early-type CENs, which becomes flatter due to containing more low 
\sersic index galaxies than our visually inspected early-type sample but 
is still steeper than the slope of S03.
There are two other possible reasons for the large
discrepancies in slope between the $r_{50}$-$L$ relation of S03 
and our result:
(1) S03 use the 
results of an early version of the NYU-VAGC fitting, which 
underestimates both the total flux 
and the size of galaxies, as shown in Appendix \ref{petro}; (2) S03 covers a wide range of luminosity and hence 
contains many faint (fainter than the faintest one in our sample) 
galaxies. 
The slope of the $r_{50}$-$L$
relation could have a smooth transition from a flat one at the faint galaxy end
 to a steep one at the luminous galaxy end, as 
suggested by \citet{desroches07}. Including many faint galaxies in the
S03's sample can flatten the average slope of the whole sample,
making their slope smaller than ours.

We also compare our 
results with other studies of the $r_{50}$-$L$ relation of BCGs, a 
subset of massive early-type galaxies at the centres of large clusters.
\citet{bernardi07} fit a \deV model to SDSS images of BCGs from the C4 
cluster catalogue \citep{miller05} and study the $r_{50}$-$L$ relation by 
using the half-light radius of their best-fitting models. They find a slope of 
$\alpha=0.89$ for early-type BCGs over the luminosity range of 
${\rm M_r-5log(h)}=[-20.3:-23.3]$. This slope is a bit shallower than our 
slope ($\sim 1.0$) for early-type CENs over the same luminosity range.
The difference could be attributed to the fact that they fit BCGs with a \deV 
profile rather than a \sersic profile, which
yields a smaller size and hence results in 
a flatter $r_{50}$-$L$ relation, as shown in their paper.
\citet{vonderlinden07} performed isophotal photometry on the SDSS BCGs from
the C4 cluster catalogue \citep{miller05} and found $\alpha=0.65\pm0.02$ for 
BCGs over the luminosity range ${\rm M_r-5log(h)}=[-20.3:-23.3]$. However, the 
isophotal photometry technique could miss the extended outer parts of galaxy 
light profiles and hence result in underestimates of both $r_{50}$ and $L$. 
\citet{lauer07a} combined surface photometry presented in several HST 
imaging programs for 219 early type galaxies and fit them using
a \deV profile.  They found $\alpha=1.18\pm0.06$ for
galaxies with ${\rm M_V}<-21$ and that the $r_{50}$-$L$
relation changed from a flat slope to a steep slope at around $M_V=-22$.
This slope is close to our slope for CENs brighter than ${\rm M_r-5log(h)}=-22$.
Also, their slope behaves qualitatively like ours, with a steep slope
at the bright end and a more shallow slope at the faint end.
\citet{gonzalez05a} fit \sersic profiles to the I-band images of 
24 luminous BCGs residing in large clusters and found $\alpha=1.8\pm0.2$, but
these fits included significant ICL. 
During our tests in \S\ref{covar}, we showed
the sensitivity to the background sky
estimate and the covariance of the size and magnitude from profile fitting.
This could become more acute for high luminosity CENs,
where distinguishing the ICL from the extended outer light
bound to a galaxy becomes increasingly difficult.

Overall, although it is 
well known that size strongly correlates with
luminosity, the slope of this relation is hard to determine precisely. 
As we summarised in Table \ref{t1}, authors with
different samples, measurement techniques, definitions of size and/or even
the criteria to split their samples into early and late types 
obtain different slopes. What's more, a small uncertainty in background sky 
translates into a strong covariance between size and luminosity and hence 
can change the slope, especially for bright and large galaxies.
Based on our analysis of the robustness and uncertainty of our fitting, and 
combined with previous work, we 
propose slopes for the $r_{50}$-$L$ relation of
$\alpha\sim0.9$ and $\alpha\sim0.6$ for early-type and late-type CENs,
respectively.

\begin{table*}
\caption {Slope of the $r_{50}$-$L$ relation for early-type galaxies 
from various authors}
\label{t1}
\begin{tabular}{cccccc}
\hline\hline
Work & Sample & $\alpha$ & $\Delta \alpha$ & Luminosity range & Model \\
\hline
This work & Y07 CENs & 1.02 & $\pm0.03$ & ${\rm M_r-5log(h)}=[-19:-24]$ & \sersic \\
This work & Y07 CENs & 0.82 & $\pm0.06$ & ${\rm M_r-5log(h)}>-22$ & \sersic \\
This work & Y07 CENs & 1.32 & $\pm0.08$ & ${\rm M_r-5log(h)}<-22$ & \sersic \\
Shen et al. (2003) & SDSS galaxies & 0.65 & -- & ${\rm M_r-5log(h)}=[-15.3:-23.3]$ & \sersic \\
von der Linden et al. (2008) & C4 BCGs & 0.65 & $\pm0.02$ & ${\rm M_r-5log(h)}=[-20.3:-23.3]$ & isophotal photometry \\
Bernardi et al. (2007) & C4 BCGs & 0.89 & -- & ${\rm M_r-5log(h)}=[-20.3:-23.8]$ & \deV\\
Lauer et al. (2006) & 219 galaxies & 1.18 & $\pm0.06$ & ${\rm M_V}<-21$ & \deV \\
Gonzalez et al. (2006) & 24 BCGs & 1.8 & $\pm0.2$ & ${\rm M_I}<-24$ & \sersic \\
\hline
\end{tabular}
\end{table*} 

Closely related to the size-luminosity relation of CEN galaxies is
their size-stellar mass $r_{50}$-\mstar relation.
For comparison, we plot the $r_{50}$-\mstar relations for our early-type
(upper right) and late-type (lower left) CENs in
Figure \ref{fig:sizebcg}. We find slopes of
$\alpha=0.90\pm0.02$ (early-type) and 
$\alpha=0.47\pm0.03$ (late-type) by using the relation
${{\rm log[}r_{50}{\rm /(h^{-1}kpc)] = \alpha log[M_{star}/(h^{-2}M_{\odot})] + \beta}}$ to fit the whole luminosity/stellar mass range of our CEN sample. 
Similar to the $r_{50}$-$L$ relations, our $r_{50}$-\mstar relations are 
steepened by the covariance between size and total magnitude in the profile
fitting. We also refit the relations with bright (${\rm M_r-5log(h)<-22}$) CENs 
excluded and find slopes of $\alpha=0.68\pm0.03$ (early-type) and
$\alpha=0.35\pm0.03$ (late-type). Our slopes of faint end 
(${\rm M_r-5log(h) \geq -22}$) 
CENs are steeper than those in S03 (thin 
dotted lines in each panel). As discussed above, differences in the
fitting procedures could be responsible
for the difference. Also we note that the $r_{50}$-\mstar relation 
in S03 is derived by using the {\it z}-band $r_{50}$.

\subsection{Comparison Between CENs and SATs}
\label{bcgsat}

One of our two primary questions regarding the environmental dependence
of morphological transformation is whether or not the
group centre is a special place for determining the structure of galaxies.
In this subsection, we compare the structural properties of CEN galaxies
with those of galaxies selected from the two SAT samples, which are comparable
to our CEN sample.
First, we consider SATs matched in stellar mass with our CEN sample, the
SAT sample S1. Second, we study SATs matched both in
stellar mass and colour, the SAT sample S2. 
(see \S\ref{sample} for details). 
We address our above question by testing if
CENs and SATs are two distinct 
populations of galaxies possessing different structural properties.
SATs were CENs before being accreted by a larger halo, thus, 
differences between these galaxies probe the impact of local environment on
SAT-specific transformation processes.

Many studies have demonstrated structural differences between
brightest cluster galaxies (BCGs) and non-BCGs. For example,
BCGs are found to have larger sizes \citep[e.g.,][]{bernardi07,vonderlinden07,liufs08}
and steeper light profiles \citep[e.g.,][]{graham96}
compared to other massive early type galaxies (ETGs)
in the same cluster. These results
are interpreted as differences between CEN and SAT galaxies and, 
as such, are thought to indicate
unique formation histories.
However, it is important to keep in mind that the BCGs in these studies
represent a special subset of all CENs; namely they are the
most-luminous and highest-mass galaxies found at the centres of massive
clusters and are the tip of the galaxy group mass function.
Moreover, a comparison of BCGs with other morphologically similar cluster 
members is effectively a comparison of galaxies of different masses.
In general, the structural properties of ETGs have a smooth
transition as their luminosity and/or stellar mass changes \citep{desroches07}.
Therefore, it is unclear whether differences between BCGs and non-BCGs
are intrinsic and originate from separate formation mechanisms or are
simply a reflection of their different masses. To avoid this selection effect,
we compare the structure of CEN and SAT galaxies that are 
matched in stellar mass.

\begin{figure}
\center{\includegraphics[scale=0.6, angle=0]{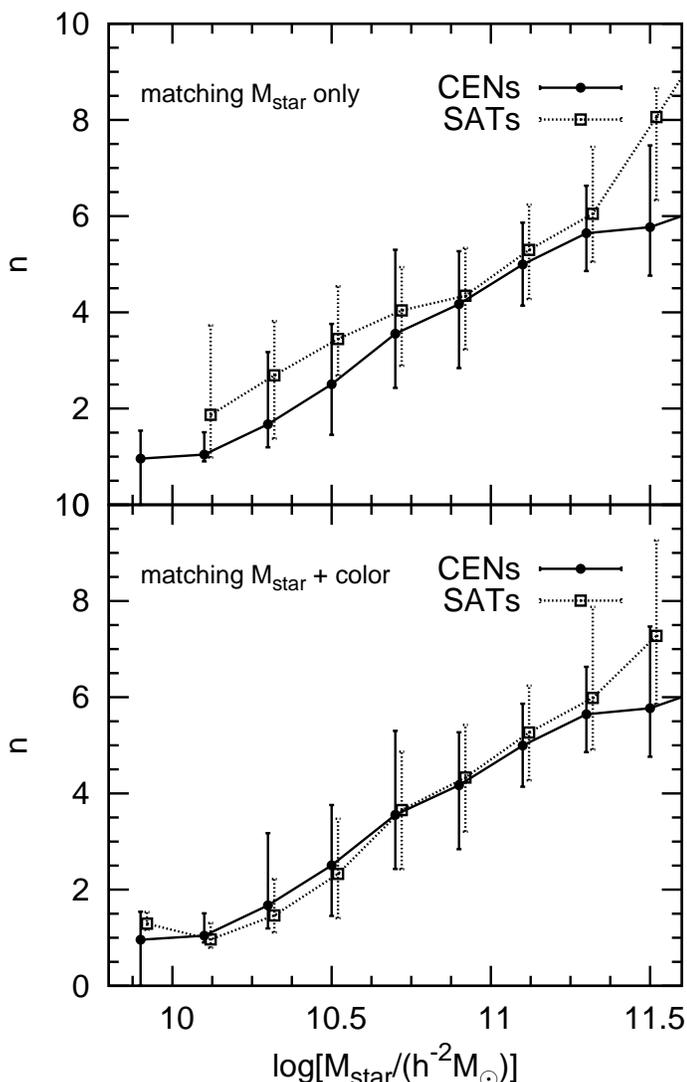}}
\caption[]{Comparison of the $n$-\mstar relation for CEN galaxies with
that of SATs matched in stellar mass
(SAT sample S1, \textit{upper} panel), and with SAT galaxies matched to
our CEN sample in both colour and stellar mass (SAT sample S2,
\textit{lower} panel). In each panel, solid lines with filled circles
represent the median values of the CEN relation
and dot-dashed lines with open squares show the SAT samples.
The errorbars give the first and third quartiles 
of the $n$ distribution in each \mstar bin as in Figure \ref{fig:nvsms}, and 
bins with less than six 
galaxies are excluded.
\label{fig:bcgvssat}}
\vspace{+0.2cm}
\end{figure}

\subsubsection{\sersic Index $n$}
\label{bcgsat_n}

We compare the $n$-\mstar relations of CENs and SATs
in Figure \ref{fig:bcgvssat}. The relation for CENs is reproduced from
Figure \ref{fig:nvsms}, and the $n$-\mstar relations of SAT sample S1 and S2
are measured in an identical fashion as for the CENS, i.e. their
stellar masses are based on their best-fitting total magnitudes from our
\sersic fitting analysis and the error bar represents the first and third
quartile in each \mstar bin. We also repeat our self-matching test 
in Sec. \ref{matching} and 
find that the scatter of \sersic index medians due to our matching 
scheme is much smaller than the sample scatter. So we decide to choose
the scatter of the sample (as shown by the first and third
quartiles in the plot) to represent the confidence interval 
whenever comparing medians of measurements from two samples. 
Note that we match our CEN and SAT samples using the
stellar mass estimated from the Petrosian magnitude in \S\ref{sample}, but
plot the result using the stellar mass estimated from the total magnitude 
in our \sersic fits. The only possible concern about this procedure
is that it may change the colour difference between CENs and SATs in 
the SAT sample S2 at given stellar mass (see the {\it lower right} panel of 
Fig. \ref{fig:matching}). To evaluate the magnitude of this effect we 
calculate the difference between the median colour of CENs and SATs in the 
SAT sample S2 in the same stellar mass bin with a width of 0.5 dex from  
${\rm log[M_{star}/(h^{-2}M_{\odot})]=9.5}$ to ${\rm log[M_{star}/(h^{-2}M_{\odot})]=12.0}$, where now the stellar mass is the one computed by using
our \sersic
total magnitude. We find a very small difference: SATs are redder by $0.015$.
However, this difference is still well within the measurement
uncertainties of colour ($\pm 0.03$) and hence our claim that the SAT sample S2
is matched with the CEN sample in both \mstar and colour is still valid. 
As for the SAT sample S1, as long as we compare CENs and SATs in same 
stellar mass bin, the modification of stellar mass does not change the results.

In the \textit{lower} panel of 
Figure \ref{fig:bcgvssat}, we see that the $n$-\mstar relation of SATs
is almost identical to that of CENs matched
in BOTH stellar mass AND colour. The only exception is for SATs in our
highest-mass bin, which have a 
somewhat higher median \sersic value than CENs of similar mass and
colour. We note that this bin has
the smallest number of SATs (13), so small number statistics may account for
the difference in the medians.
However, if we release the constraint on colour 
matching, low mass SATs (${\rm log[M_{star}/(h^{-2}M_{\odot})]<10.75}$) tend 
to have a somewhat higher median \sersic index than 
their CEN counterparts (\textit{upper} panel), although the discrepancy
is within the scatter. In contrast,
the more massive SATs have similar $n$-\mstar relations as CENs matched only in
stellar mass, except again for the highest-mass SATs.
Our results are in good agreement with 
those obtained by \citet{vandenbosch08a}, who used concentration defined
by the ratio of SDSS radii containing
90 percent and 50 percent of the Petrosian flux, rather than $n$, to describe the
profile shape
of galaxies and found that low mass SATs are redder and 
slightly more concentrated if matched with similar mass CENs.
In addition, \citet{vandenbosch08a} found that the concentration 
difference goes to zero when matched in colour and M as we find here.

\subsubsection{Size}
\label{bcgsat_size}

\begin{figure*}
\center{\includegraphics[scale=0.8, angle=0]{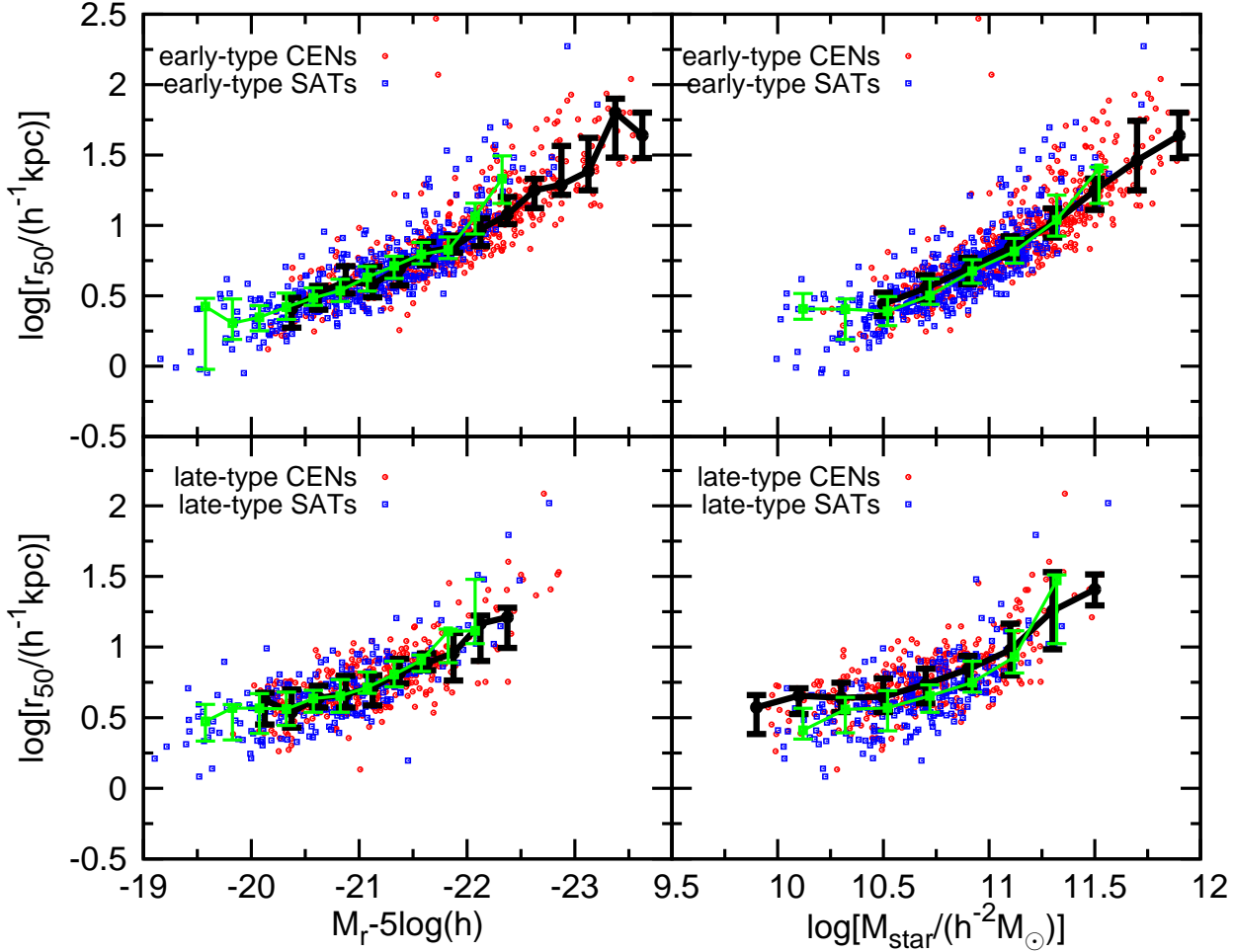}}
\caption{Comparison of CEN (red points) and SAT (blue
points) galaxy size-luminosity ({\it left} column)
and size-stellar mass ({\it right} column) scaling relations.
These SATs (sample S1) have similar stellar masses as our CEN sample.
The black (green) lines and errorbars show the median, first and 
third quartiles of the CEN (SAT) distribution 
in each luminosity or stellar mass bin. Both CENs and SATs are separated
into early-type ({\it upper} panels) and late-type ({\it lower} panels) 
galaxies (see text for detail).
\label{fig:sizecomp5}}
\end{figure*}

\begin{figure*}
\center{\includegraphics[scale=0.8, angle=0]{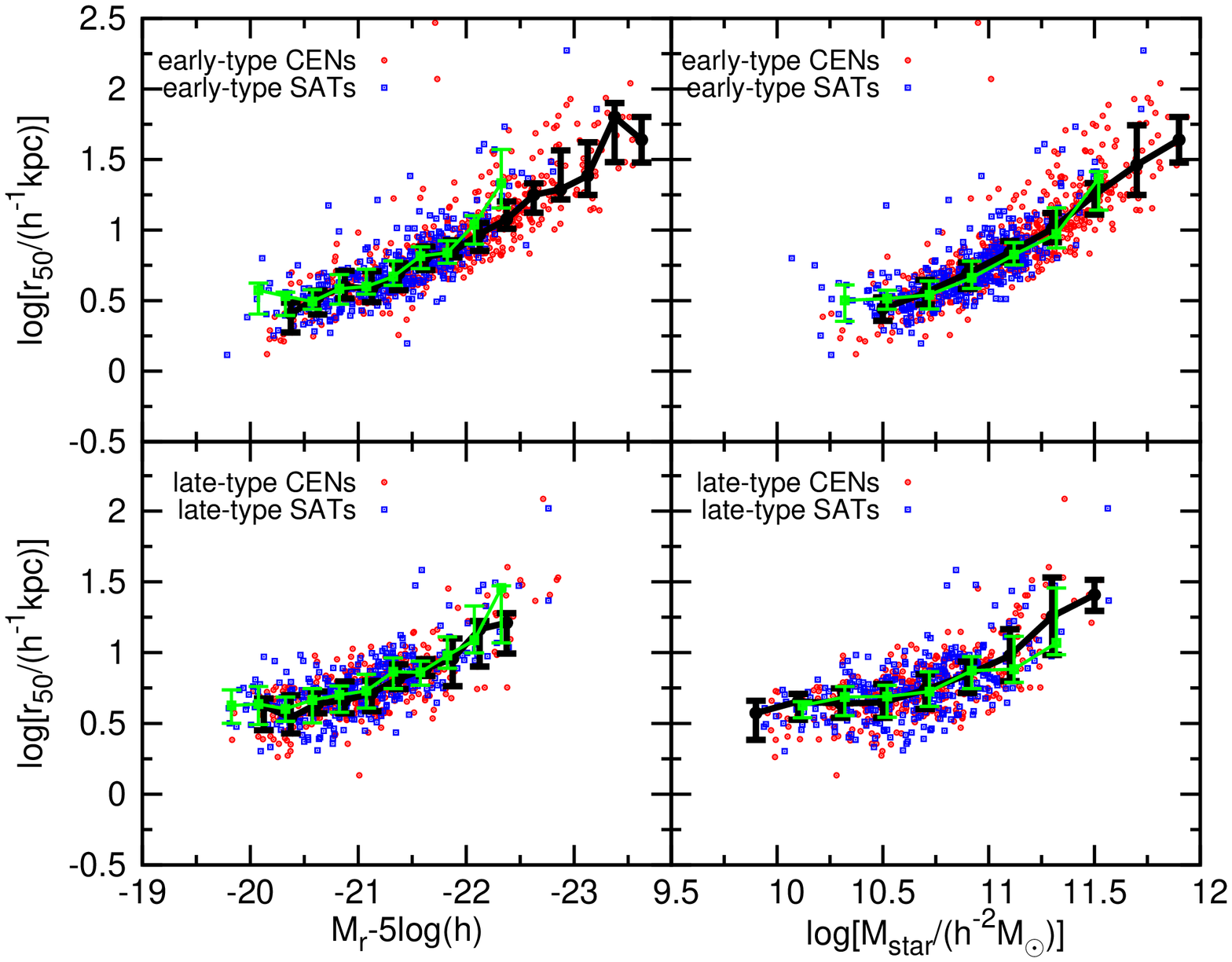}}
\caption{Same as Figure \ref{fig:sizecomp5}, except comparing CENs
with SATs of similar stellar mass and colour. 
\label{fig:sizecomp}}
\end{figure*}

Besides the $n$-\mstar relations, we use
the $r_{50}$-$L$ and $r_{50}$-\mstar relations to compare the 
structural properties of CEN and SAT galaxies.
We plot the sizes, luminosities, and stellar masses of our CEN sample
from Figure \ref{fig:sizebcg} as red points in Figures \ref{fig:sizecomp5} 
and Figure \ref{fig:sizecomp}.  The blue points represent
the data for our SAT samples S1 (Fig. \ref{fig:sizecomp5}) and S2 
(Fig. \ref{fig:sizecomp}).
The size, absolute magnitude and stellar mass of SATs are calculated in 
the same way as CENs as described in \S\ref{sersic} and \S\ref{size}.
Also, as in \S\ref{size}, we divide the SAT samples S1 and S2 into
early-type and late-type galaxies based on our visual inspection. 
However, instead of computing the slopes of
the CEN and SAT scaling relations, we
compare the half-light size distributions (median, first, and third quartiles)
of each sample in narrow bins of luminosity (0.25 mag wide for the
$r_{50}$-$L$ relation) and stellar mass (0.2 dex wide for $r_{50}$-\mstar). 
In this manner, we
directly compare CENs with SATs from each sample only in regions of overlapping
luminosity or stellar mass. Hence, we avoid comparing scaling relation
slopes based upon samples that span different luminosity
and stellar mass ranges.  This helps to avoid the effects of
small number statistics at the bright/massive and the faint/low-mass
ends, which can bias the scaling relation slopes.

The comparisons
in Figures \ref{fig:sizecomp5} and \ref{fig:sizecomp} show that for
early type galaxies, whether or not they are matched in colour, CENs and
SATs display almost no difference in their size distributions.
The only differences occur at
the bright/massive ends of the SAT relations, where both 
SAT sample S1 and S2 suffer from 
small number statistics. For late-type SAT galaxies, the two samples
also have size-luminosity relations that are similar to the CENs (the 
\textit{lower left} panel in both figures). SATs have
smaller median sizes than CENs when we only match in stellar mass (the
\textit{lower right} panel of Figure \ref{fig:sizecomp5}), but 
this difference disappears when we compare to SATs matched in both colour
and stellar mass (the \textit{lower right} panel of Figure \ref{fig:sizecomp}). 

\subsubsection{Discussion}
\label{bcgsatdiscuss}

Using the
$n$-${\rm M_{star}}$, $r_{50}$-$L$ and $r_{50}$-\mstar
relations, we compare galaxies in our CEN sample to SATs of the same
stellar mass. We find two basic differences between CEN and SAT
galaxies:
(1) low mass (${\rm log[M_{star}/(h^{-2}M_{\odot})]<10.75}$) SATs have a
slightly higher median \sersic index compared to CENs of the same stellar mass;
and (2) low-mass, late-type 
SATs have smaller median sizes compared to their same-mass CEN counterparts. 
Our findings are in good qualitative 
agreement with the results of \citet{weinmann08} for a much larger SDSS sample.
Using the NYU-VAGC half-light radii and concentrations for over $10^5$
galaxies, they found that 
late-type SATs are smaller and more concentrated than late-type CENs of the
same stellar mass. We note, however, that their size measurements suffer from
a systematic
underestimate owing to an overestimate of the background sky levels 
(as we discuss in Appendix \ref{petro}). 

We also find that the above two basic differences
disappear when we compare CENs and SATs of the
same stellar mass {\it and optical colour}. Moreover, we find no structural
differences between high-mass
(${\rm log[M_{star}/(h^{-2}M_{\odot})]>10.75}$) CEN and SAT galaxies,
which tend to have early-type
(spheroid-dominated) morphologies, or between early-type CENs and SATs,
in general. 

For lower-mass (${\rm log[M_{star}/(h^{-2}M_{\odot})]<10.75}$) 
SATs, the minor differences with CENs of similar stellar mass, and the lack
of any difference with CENs of the same mass and colour, can be understood in
terms of our selection criteria for the SAT sample S1 and S2.
According to the hierarchical
scenario of structure formation, larger haloes housing groups and clusters
are formed through the accretion and merging of smaller haloes. Thus, all
SATs were once the CEN galaxy in a smaller halo. 
Under this assumption,
low-mass SATs tend to be redder on average than CENs of similar stellar mass
because SATs have had their star formation quenched by environmental processes
once they became non-CEN members of a larger halo \citep{vandenbosch08a}.
In addition,
as Weinmann et al. (2008) point out, quenching will also cause a moderate
increase in the concentration and decrease in the size of late-type SATs,
as we find here when comparing similar mass SATs and CENs.
It is important to keep in mind that the average 
structural differences are small
and, as such, 
SAT quenching cannot be used to explain the major
morphological transformation of many disks into spheroids that is
required
to produce the strong morphological bimodality between blue and red
galaxies.
When we restrict our SAT selection to match lower-mass CENs in {\it both} 
stellar mass and colour (SAT sample S2),
we preferentially choose blue SATs.
Lower-mass CENs are typically blue
(Figure \ref{fig:samplecheck}) and
their colour is associated with late-type morphology (disk-like) and
on-going star formation. Therefore, we argue that blue SATs are likely examples
of newly accreted SATs, i.e. recent CENs of the
most recently accreted subhaloes. If true, new SATs have not been
members of larger haloes long enough to alter their colour and structural
properties. This line of reasoning agrees with the absence of 
structural differences we find for the CENs and SATs in SAT sample S2. 

When we consider higher-mass 
(${\rm log[M_{star}/(h^{-2}M_{\odot})]>10.75}$) SATs (either sample S1 or S2),
they have the same red colour (see 
lower left panel of Figure. \ref{fig:matching}) and highly-concentrated,
large \sersic index profiles associated with early-type
morphologies as CENs of similar mass.
While massive SATs are rare compared to their CEN counterparts, they are
structurally indistinguishable. In a broader sense, we find that all
morphologically early-type SATs and CENs have the same size scaling relations
and that any reported differences between CENs and SATs
\citep[e.g.][]{graham96,liufs08} are actually
the result of comparing two populations, e.g. BCGs and non-BCGs, with different
stellar mass distributions.
Owing to the similar red colours of massive CENs and SATs, 
there is no way
to discern recent arrivals from long-term members in larger haloes,
but it is clear that the transformation into spheroids
does not depend on becoming a SAT.  If that were true, we would expect
massive CENs to be disk-like when they are clearly otherwise.
Rather, we argue that the strong morphological
transformation from disk to spheroid
occurred at an earlier time when a massive SAT was the CEN of a smaller
halo and that the local environment had no additional
impact on the structure of high-mass spheroids.
We find that the difference between disk-dominated and spheroid-dominated
structure is more directly related to the stellar mass of a galaxy.
Clearly, there is some relationship between the mass of CENs and their host
halo mass (Figure \ref{fig:samplecheck}), but further study is required
to understand whether or not any aspect of the environment plays an
important role in the transformation of
disks and the production of high-mass spheroids.

\section{Summary}
\label{summary}

In this paper, we study how the structural properties of central galaxies 
(CENs) in groups and clusters depend on galaxy stellar mass,
global environment (group halo mass), and
local environment (central/satellite position within the host halo).
We select
from the SDSS DR4 group catalogue \citep{yang07}
a statistically representative sample of 911 CENs 
whose host halo masses span from
$10^{12}$ to ${\rm 10^{15}h^{-1}M_{\odot}}$. We use 2D \sersic model fits
to quantify the shape (\sersic index) and size (half-light radius) of each
galaxy. To this end, we establish a well-tested, GALFIT-based pipeline to fit 
\sersic models to SDSS imaging data in the $r$-band.
We summarise our main findings below.

We thoroughly test the performance of our GALFIT pipeline on simulated and real
SDSS galaxy image data. Our 2D fitting recovers the structural properties
of simulated galaxies with no bias,
unlike the one-dimensional fits to 
azimuthally-averaged data employed for the NYU-VAGC that 
systematically underestimate
the total flux, size and \sersic index of higher-$n$ profiles. 
For galaxy profile fitting, we also demonstrate that 
the SDSS global sky is preferred over the SDSS local value as a background 
level measurement. We compare our fitting results with those
from the NYU-VAGC and find that our fits include light from the outer parts 
of galaxies, which is missed when an overestimate of the (local) sky background 
is used.
We test how this background uncertainty
translates into a systematic uncertainty in the fitting parameters owing to a
strong covariance between \sersic index, total
magnitude, and half-light size. This 
covariance affects bright and large galaxies more and could contribute to the
apparent steeping in the slope
of the size-luminosity and size-stellar mass relations at the bright
(massive) end.

We find that the \sersic index of CENs depends strongly on ${\rm M_{star}}$, 
but weakly or not
at all on ${\rm M_{halo}}$. The dependence on stellar mass is
in the sense that low mass galaxies (${\rm M_{star}<10^{10.5}h^{-2}M_{\odot}}$) have 
lower, disk-like indices ($n\sim 2.0$), while 
massive galaxies (${\rm M_{star}>10^{11.0}h^{-2}M_{\odot}}$) have higher,
spheroid-like
($\sim 5$) indices. 
Over a large range in \mhalo, from small groups to large clusters,
any change in the $n$ distribution of CENs is likely the result of the
correlation between
\mstar and ${\rm M_{halo}}$. The fact that spheroidal CENs are found at all 
group masses, and the lack of a strong $n$ dependence on \mhalo, both rule out
a distinct halo mass for producing spheroids. 
Moreover, the strong dependence of $n$ on 
\mstar suggests that \mstar is the key factor in determining the shape of CENs.

Similar to the light profile shape, the half-light size of CENs depends on 
galaxy stellar mass and luminosity.  
We separate our CEN sample into early and late-type galaxies by 
visual inspection and we 
find a $r_{50}$-$L$ slope of 
$\alpha \sim 0.83 (0.62)$ for early-type (late-type) galaxies 
with $-22<{\rm M_r-5log(h)}<-20$.
We also compare our 
$r_{50}$-$L$ slope for early-type CENs with those from other studies
and find that there is fairly large discrepancy.
This discrepancy could result from several factors 
including different samples, size measurement
techniques, or early-type galaxy definitions.

To study whether the structural properties of CENs depend on their
special position at the centre of the gravitational potential well, 
we compare their shapes and sizes with those of non-CEN satellite (SAT)
galaxies. 
We find that low mass ($<10.0^{10.75}{\rm h^{-2}M_{\odot}}$) SATs have somewhat
larger median \sersic indices compared with CENs of similar stellar mass.
In addition, low mass late-type SATs are moderately smaller in size than
late-type CENs when matched in stellar mass, but no size differences are found
between early-type CENs and SATs.
We find {\it no structural differences} between SATs and CENs when they are 
{\it matched in both optical colour and stellar mass.}
The small differences in the sizes of low-mass, late-type CENs and SATs
are consistent with SAT quenching as found by others
(e.g., in \citet{vandenbosch08a} and \citet{weinmann08}).
The similarity in the structure of massive SATs and massive CENs
demonstrates that 
the local environment has no significant impact
on the structure of a massive galaxy that enters a denser environment
and that these two populations are morphologically indistinguishable.

We conclude that \mstar is the most fundamental property in determining the
basic structural shape and size of a galaxy.
In contrast, the lack of a significant $n$-\mhalo relation rules out
a clear distinct group mass for producing spheroids. This fact, combined with
the existence of spheroid CENs in low-mass and high-mass groups, suggests
that the strong morphological transformation processes that produce
spheroids must occur at the centres of groups spanning a wide range of masses.

\section*{Acknowledgments}
We made extensive use of the SDSS SkyServer Tools
({\texttt http://cas.sdss.org/astro/en/tools/}).
We appreciate useful discussions with Marco Barden, Boris H{\"a}ussler and
Chien Peng.
We thank Micheal Blanton for providing the SDSS simulations and for
useful discussions.
D.\ H.\ M.\ and N.\ K.\
acknowledge support from the National Aeronautics
and Space Administration (NASA) under LTSA Grant NAG5-13102
issued through the Office of Space Science.
Funding for the SDSS
has been provided by the Alfred P.\ Sloan Foundation, the
Participating Institutions, the National Aeronautics and Space Administration,
the National Science Foundation, the U.S. Department of Energy,
the Japanese Monbukagakusho, and the Max Planck Society.  The SDSS
Web site is {\texttt http://www.sdss.org/}.  The SDSS is managed by the
Astrophysical Research Consortium (ARC) for the Participating Institutions,
which are The University of Chicago, Fermilab, the Institute
for Advanced Study, the Japan Participation Group, The Johns Hopkins
University, Los Alamos National Laboratory,
the Max-Planck-Institute for Astronomy (MPIA), the Max-Planck-Institute for
Astrophysics (MPA), New Mexico State University, University of Pittsburgh,
Princeton University, the United States Naval Observatory, and
the University of Washington.
This publication also made use of NASA's Astrophysics Data System
Bibliographic Services.

\bibliographystyle{mn2e}
\bibliography{references}

\appendix

\section[]{Comparison with NYU-VAGC \sersic Fits}
\label{petro}

\begin{figure*}
\center{\includegraphics[scale=0.9, angle=0]{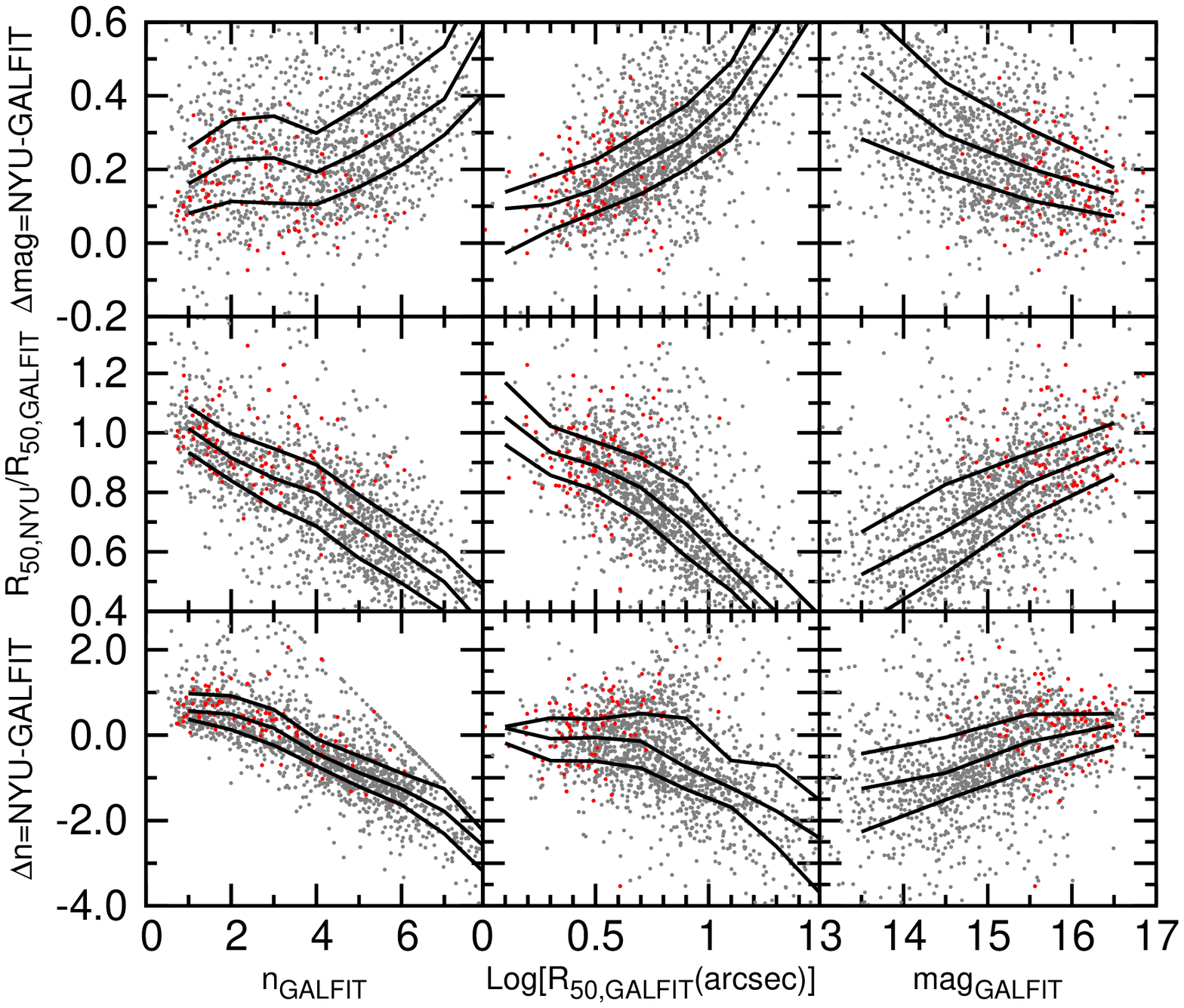}}
\caption[]{Differences between the NYU-VAGC and GALFIT \sersic 
parameters from $r$-band SDSS data as a function of the GALFIT
\sersic parameters for our total sample of 1657 CEN+SAT galaxies. 
The lines show the quartiles of each distribution as in 
Fig. \ref{fig:simmatrix}. Black points are for galaxies with
$|\mathrm{\Delta sky|>0.1\ ADU}$ and
red points for $|\mathrm{\Delta sky|<0.1\ ADU}$, 
where $\mathrm{\Delta sky=local\ sky - global\ sky}$.
Note, in the \textit{left} panels
it is clear that the NYU-VAGC fitting has an $n=6$ limit.
\label{fig:matrix_quart}}
\vspace{-0.2cm}
\end{figure*}

\citet[][hereafter 'NYU-VAGC']{blanton05nyu} fit \sersic models to the 
azimuthally-averaged 1D profiles output by the SDSS photometric pipeline 
\citep{stoughton02} for 
all SDSS DR4 galaxies meeting the Main sample criteria.
Tests show 
that the NYU-VAGC \sersic fitting does well for simulated galaxies with 
an input \sersic index $n_{\rm in}<2$, a small size or a faint 
magnitude. But for $n_{\rm in}>2$ simulations,
the NYU-VAGC fitting systematically underestimates $n$, $r_{50}$, and 
total flux [see Figure 9 in \citet{blanton05nyu} for 
details]. For example, for a simulated galaxy with $n_{in}=4$ the 
NYU-VAGC fitting underestimates
these parameters by 15 percent. For comparison, our GALFIT fitting results for 
850 simulated \sersic galaxies placed in SDSS images show very little 
bias for $n_{in}>4$ galaxies (see Figure \ref{fig:simmatrix}). 
Note that we adopt the global sky value from the SDSS
image header in our GALFIT profile fits, while NYU-VAGC uses the
local sky level. It has been reported that the SDSS pipeline overestimates
the local sky in dense environments \citep{lauer07a,vonderlinden07,adelman08}. 
As such, \sersic
fits based on overestimates of the sky may result in fainter magnitudes, 
smaller sizes and/or 
lower \sersic indices as we demonstrate in \S\ref{covar}.

Figure \ref{fig:matrix_quart} shows comparisons between our fit results and
those from NYU-VAGC for our total sample of 911 CENs plus 746 SATs from our SAT 
sample S2 (see \S\ref{sample} for details). 
Below, we discuss in detail the discrepancies between the two 
fits for galaxies with $n>3$ from GALFIT, 
and then for those with lower \sersic indices.

\subsection{High \sersic Galaxies}
\begin{figure}
\center{\includegraphics[scale=0.5, angle=0]{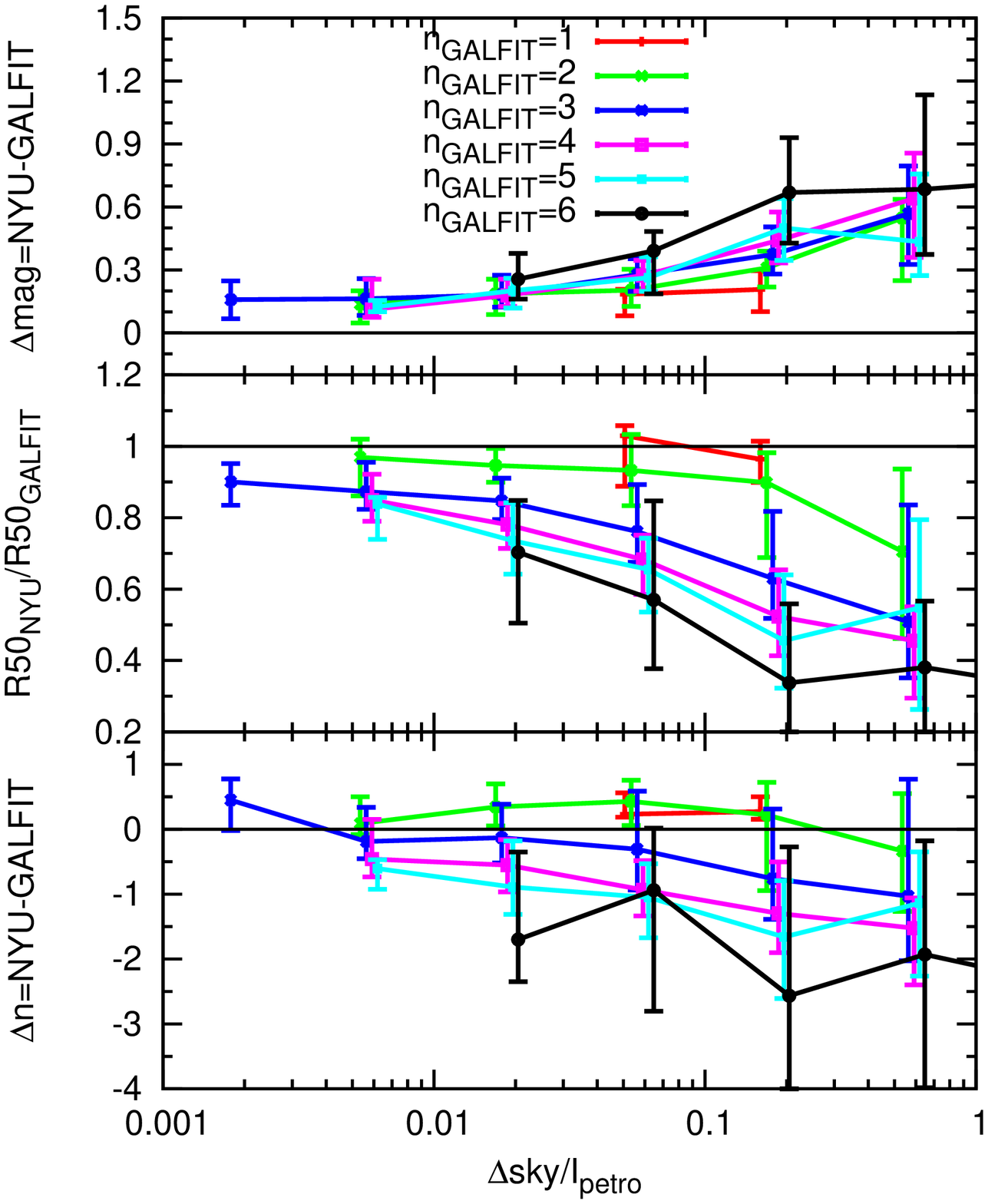}}
\caption{The NYU-GALFIT discrepancies shown in 
Figure \ref{fig:matrix_quart} as a
function of sky difference, divided into GALFIT \sersic index bins as
shown by the colour coding. Here the 
sky difference is expressed by the ratio between
$\mathrm{\Delta sky=local\ sky - global\  sky}$
and the average Petrosian surface brightness
(Petrosian quantities are directly drawn from SDSS tables).
\label{fig:dn_sky}}
\end{figure}

For $n>3$ galaxies, the NYU-GALFIT parameter discrepancy for real galaxies 
in Figure \ref{fig:matrix_quart}
follows a similar trend as those between the input and fit parameters for simulated galaxies
in Figure 9 of \citet{blanton05nyu}, but with an increased amplitude. 
For example, the \sersic difference (lower left panel) grows by about 
$\Delta n \simeq 1.3$ over the interval $3<n<6$, compared with 
$\Delta n \simeq 0.6$ over the same interval in the \citet{blanton05nyu} 
simulations.
Here two factors are at play: 
one is the systematic underestimate of NYU-VAGC's 1D fitting procedure for 
steep \sersic profiles as demonstrated in their test fitting using
simulations.
The second factor is the difference between the sky levels used in each 
procedure.
In Figure \ref{fig:dn_sky}, we attempt to separate these two factors by 
splitting the whole sample into GALFIT
\sersic index and relative sky difference bins,
where the relative sky difference 
is the difference between the local and the global
sky normalised by the Petrosian surface brightness of the galaxies
($I_{petro} = f_{Petro}/(2 \pi r^2_{50,Petro})$ in units of ADUs 
per pixel, where $f_{Petro}$ and $r_{50,Petro}$ are the Petrosian flux and 
Petrosian half-light radius, respectively). For $n>4$ galaxies 
with a normalised 
sky difference less than 0.01, meaning that the sky difference is at
most a minor issue, the $n$ and $r_{50}$ disagreements
more or less reflect the systematic underestimates seen in the NYU-VAGC 
fitting of simulated galaxies. As $\Delta sky/I_{petro}$ increases, the 
NYU-GALFIT disagreements grow and we see a trend of larger NYU-VAGC 
underestimates
for galaxies with higher $n$, as expected when the
local sky estimate includes more of the light belonging to each galaxy.

\subsection{Low \sersic Galaxies}
We have outlined how the NYU-VAGC and GALFIT methods both do very well in fitting
pure-\sersic simulations with
$n_{\rm in}<3$,
therefore we expect minor differences when comparing fits to real galaxies with 
disk-like profiles. However, we find that the fit parameters from the two procedures 
differ in two ways for $n<3$ galaxies, as shown in 
Figure \ref{fig:matrix_quart}. First, the NYU-VAGC fits have systematically 
higher \sersic values
than the GALFIT fits, which is inconsistent with the results from the simulations. Second, there is a systematic offset of about 0.2 mag between
the magnitude of NYU-VAGC fits and our GALFIT fits
in the sense that NYU-VAGC finds fainter fluxes. We note that the
offset appears to be independent of the difference in the sky ($\mathrm{\Delta sky=local\ sky\ -\ global\ sky}$) used in each fitting procedure, as shown by the 
similarity between the red (small $\mathrm{\Delta sky}$) and the black (large $\mathrm{\Delta sky}$)
points for $n<3$ galaxies in Figure \ref{fig:matrix_quart}.

We suspect that the NYU-VAGC procedure of fitting a 1-D \sersic model to azimuthally averaged annuli overestimates \sersic indices for disk 
galaxies. We check the distribution of \sersic indices for the whole 
NYU-VAGC and find that the number of galaxies with $0.5<n_{\rm NYU-VAGC}<1.0$ is
much less than those with $1.0<n_{\rm NYU-VAGC}<1.5$. This results in 
conflicts with other observations of disk galaxies
\citep[e.g.][]{driver06,vanderwel08,haussler07}.
To test our suspicion, we visually inspect galaxies from our sample
with a late-type fit ($n<2$) by GALFIT, 
but an early-type fit ($n_{\rm NYU-VAGC}>2.5$) by NYU-VAGC.
To exclude the sky influence, we restrict our inspection to 36 galaxies with
$|\Delta {\rm sky}|<0.5 {\rm\ ADU}$. At least two thirds of these galaxies
have very obvious spiral features as expected for galaxies with
disk-dominated light profiles. Another 20 percent have disturbed morphologies
or very bright nearby stars, which could cause spurious fits. 
A majority of the spirals are inclined with $b/a<0.5$. 
As clearly demonstrated by \citet{bailin08}, $n_{NYU-VAGC}$ is systematically 
overestimated for more inclined galaxies. This effect is the result of 
edge-on or inclined galaxies having steeper
azimuthally averaged radial profiles
because the averaged flux from the narrow outer part of such 
galaxy is decreased by being smoothed over a large circular area. 

\subsection{Comparing \sersic Magnitude Estimates}

\begin{figure}
\center{\includegraphics[scale=0.5, angle=0]{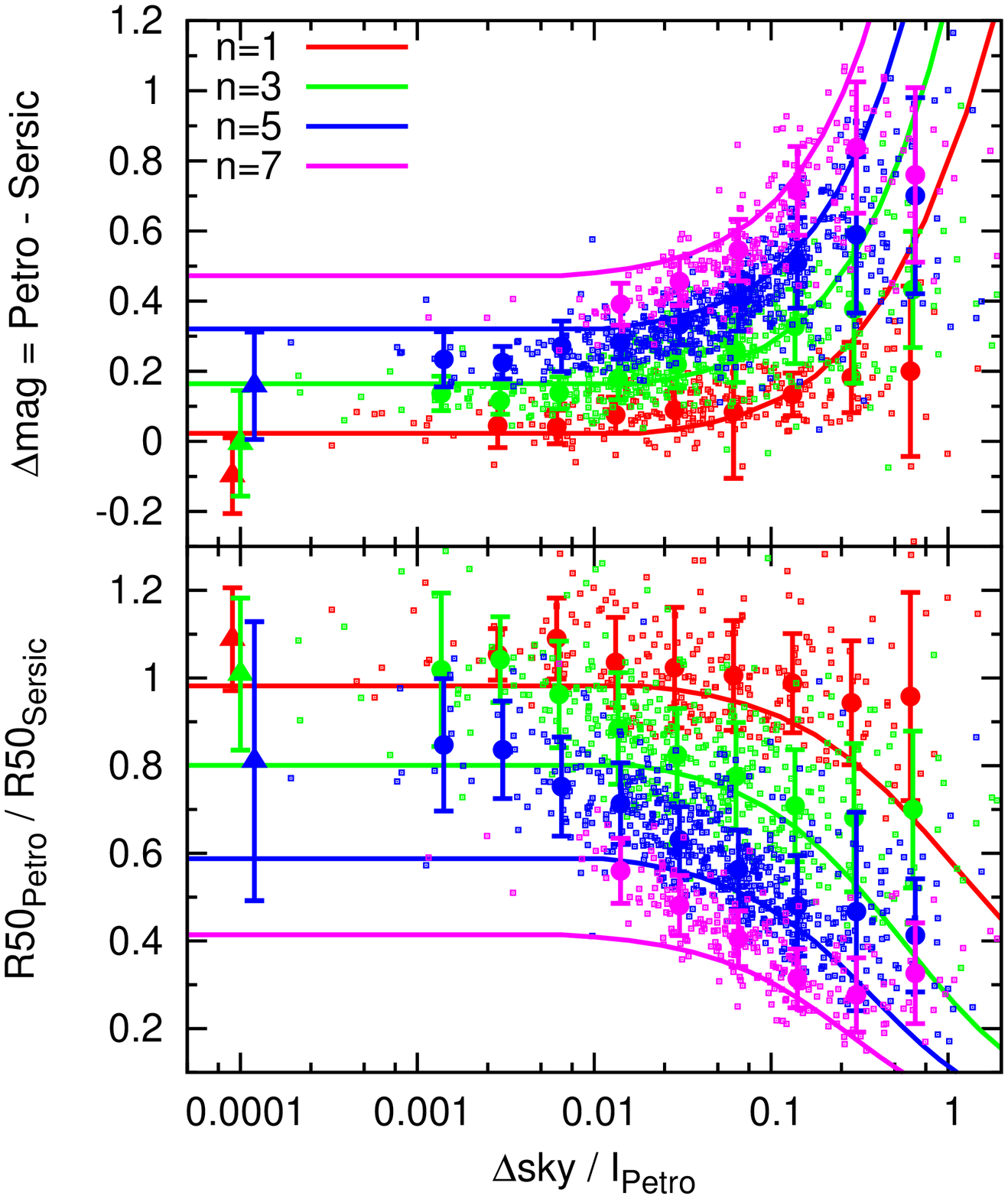}}
\caption{The difference between Petrosian and \sersic model quantities 
as a function of sky offset normalised to the Petrosian surface 
brightness.
Cases with different 
\sersic index are colour coded as indicated in the top panel. Solid lines 
are for our predictions based 
on a pure \sersic model (see text for details), for which $\Delta {\rm sky}$ 
means the imaginary overestimation of sky background. The small squares are for each galaxy in our GALFIT sample, for which $\Delta {\rm sky}$ means the SDSS
local-global sky difference. The filled circles with error bars show the mean and standard deviation of our sample 
in different bins. We also compare with the mean and standard deviation 
(triangles with error bars) of the
NYU-VAGC fitting parameters,
which should be plotted at $\mathrm{\Delta sky =0}$, but are shifted a little to 
allow them to be plotted on a log-scale plot.
\label{fig:petro}}
\end{figure}

Besides the tendency to overestimate the \sersic indices of actual disk-dominated
galaxies when using 1D fits to azimuthally-averaged radial profiles, 
we also explore the offset between the NYU-VAGC and the GALFIT \sersic magnitudes in more detail.
For this exercise we use an independent measure of galaxy flux, the
SDSS Petrosian magnitude, to anchor our comparisons of different \sersic
magnitudes from the two methods.
The SDSS photometric pipeline calculates the flux within a circular 
aperture equal to two times the Petrosian radius, which provides an
approximate total galaxy magnitude. It is well known, however, that
Petrosian magnitudes systematically miss some flux when applied to
different \sersic model profiles.
As shown in \citet{graham05}, the Petrosian magnitude
misses very little flux for an $n=1$ profile, but for an $n=4$ galaxy
it will underestimate the brightness
by about 0.2 magnitudes. We note that the \citet{graham05} calculations are
valid only when the sky is known perfectly.
Any under/overestimation of the sky background will increase/decrease the
discrepancy between the Petrosian and the \sersic magnitudes. Using the formalism of
\citet{graham05}, we predict the Petrosian-\sersic magnitude offset
($\Delta{\rm mag}={\rm Petrosian - Sersic}$)
under the influence of different amounts of overestimation of the real sky
 by subtracting a range of background
pedestals to each \sersic model before measuring the Petrosian flux.
Our $\Delta{\rm mag}$
predictions for different $n$ are shown
in Figure \ref{fig:petro} (top panel) as a
function of the sky overestimate ($\Delta {\rm sky}$), 
normalised by the Petrosian surface brightness (as in Figure \ref{fig:dn_sky}).
When the normalised sky difference is less than $10^{-4}$, our predictions
converge as expected to the values claimed in \citet{graham05}. However, as
the overestimates of the local sky increase,
underestimates of the Petrosian magnitude for different $n$ grow systematically.
Likewise, we also make predictions for the offsets between the Petrosian
and the \sersic half-light radii ($r_{50,{\rm Petro}}/r_{50,{\rm Sersic}}$)
as a function of sky offsets and plot these in the bottom panel of
Figure \ref{fig:petro}.

In Figure \ref{fig:petro}, we also compare our GALFIT results for actual 
galaxies (small squares)
to the sky dependent predictions. Our working assumptions
are: (1) a \sersic model is
a reasonable model to describe galaxy light profiles, and (2)
the SDSS global sky is a good measurement
of the real sky and is preferred to using the SDSS local value. 
Here $\Delta {\rm sky}$ is the local-global sky
difference in SDSS and all the Petrosian 
results are measured using the local sky.
In the {\it upper} panel, we see that our fit results are close to the 
$\Delta{\rm mag}$ predictions for a wide range
of sky differences and \sersic indices.
We also find fair agreement between our $r_{50}$ results and the 
predictions in the {\it lower} panel of
Figure \ref{fig:petro},
suggesting that our fitting results are self-consistent under the
two assumptions above.

Finally, we compare the NYU-VAGC results for the real data with the predictions.
Given that the NYU-VAGC fitting uses the same local sky as the Petrosian quantities, 
all the galaxies with NYU-VAGC fits have $\Delta{\rm sky}=0$ by
definition. Therefore, the values of $\Delta{\rm mag}$ and
$r_{50,{\rm Petro}}/r_{50,{\rm Sersic}}$
for the NYU-VAGC \sersic results for our sample (triangles in Figure \ref{fig:petro})
should satisfy the predictions of \citet{graham05}. Yet, we see
that the NYU-VAGC \sersic results are actually underestimates, on average,
compared to the predictions. For example, the $n=1$ galaxies 
have \sersic magnitudes that are
0.1 mag {\it fainter} than the Petrosian measurement, which is 
{\it inconsistent}
with either the predictions
or the definition of the two magnitudes.
By definition, \sersic magnitudes are based on a model flux integrated
to infinity, thus there is no reason for such a magnitude to be fainter
than the Petrosian {\it aperture} magnitude, which only includes 
light our to some
radius. It is unclear why the NYU-VAGC \sersic fitting procedure produces fainter
magnitudes than expected, but this effect combined with the nonzero 
$\Delta {\rm sky}$ values explain the systematic
0.2 mag offset that
we find between the GALFIT and the NYU-VAGC \sersic magnitudes for $n<3$ galaxies
(Figure \ref{fig:matrix_quart}).
Based on the above analysis, we conclude
that if the two assumptions of our fitting are valid, i.e. assuming a \sersic
model is the correct model and that the SDSS global sky is an accurate measure
of the true sky background, then our GALFIT
fitting of the SDSS data returns more accurate measurements for the
structural parameters of galaxies than those in the NYU-VAGC.

\bsp

\label{lastpage}

\end{document}

%% file: macros.tex
\newcommand{\sersic}{{S\'{e}rsic }}

\newcommand\mhalo{${\rm M_{halo}}$ }
\newcommand\mstar{${\rm M_{star}}$ }
\newcommand\deV{de Vaucouleurs }

\newdimen\hssize 
\newdimen\hdsize 